\documentclass[10pt,prd,a4paper,onecolumn,nofootinbib,superscriptaddress,floatfix,preprintnumbers,eqsecnum]{revtex4}

\usepackage{graphicx,multirow}
\usepackage{caption}
\usepackage{subcaption}
\usepackage{dcolumn}
\usepackage{bm}
\usepackage[all]{xy}
\usepackage{color}
\usepackage{amsmath,amssymb}
\usepackage[usenames,dvipsnames,svgnames]{xcolor}
\usepackage[unicode=true,pdfusetitle,
bookmarks=true,bookmarksnumbered=false,bookmarksopen=false,citecolor=DodgerBlue,
breaklinks=false,pdfborder={0 0 1},backref=false,colorlinks=true,linkcolor=DodgerBlue,pdfpagemode=FullScreen]
{hyperref}
\usepackage[utf8]{inputenc}

\usepackage{slashed}
\usepackage{physics}
\usepackage{varwidth}
\usepackage{braket}

\newcommand{\GeV}{\textrm{ GeV}}
\newcommand{\TeV}{\textrm{ TeV}}
\makeatletter

\newcommand{\beq}{\begin{equation}}
\newcommand{\eeq}{\end{equation}}
\newcommand{\bea}{\begin{eqnarray}}
\newcommand{\eea}{\end{eqnarray}}

\newcommand{\Oone}{\mathcal{O}_{1}}
\newcommand{\Ofour}{\mathcal{O}_{4}}
\newcommand{\Oten}{\mathcal{O}_{10}}
\newcommand{\Oeleven}{\mathcal{O}_{11}}
\newcommand{\lsim}{\:\raisebox{-0.5ex}{$\stackrel{\textstyle<}{\sim}$}\:}
\newcommand{\gsim}{\:\raisebox{-0.5ex}{$\stackrel{\textstyle>}{\sim}$}\:}

\usepackage{color}

\DeclareRobustCommand{\rchi}{{\mathpalette\irchi\relax}}
\newcommand{\irchi}[2]{\raisebox{\depth}{$#1\chi$}} 

\begin{document}
\title{Neutron EDM constrains direct dark matter detection prospects}

\author{Manuel Drees}
\email{drees@th.physik.uni-bonn.de}
\author{Rahul Mehra}
\email{rmehra@th.physik.uni-bonn.de}
\affiliation{Bethe Center for Theoretical Physics and Physikalisches Institut, Universit\"{a}t Bonn, Nussallee 12, D-53115 Bonn, Germany}

\begin{abstract}
		
  A non-relativistic effective field theory (NREFT) offers a bottom-up
  framework to classify Dark Matter (DM) -- nucleon interactions
  relevant for scattering at direct detection experiments by
  organizing the interactions in powers of the momentum transfer
  $ \vec{q} $ and DM velocity $ \vec{v} $. This approach generates a
  number of operators including P-odd and T-odd operators; these can
  only be generated from a relativistic theory with CP violating
  interactions. We consider the leading order P-odd, T-odd operators
  viz.  $ \mathcal{O}_{10} $, $ \mathcal{O}_{11} $ and
  $ \mathcal{O}_{12} $ and compare the constraints on these operators
  from leading direct detection searches and from the bound on the
  neutron EDM (nEDM). We perform our analysis using simplified models
  with charged mediators and compute the loop diagrams contributing to
  the nEDM. We find that constraints on the DM scattering cross section
  from the bound on the nEDM are several orders of magnitude stronger
  than the limits from direct searches, and even well below the
  neutrino floor for such NREFT operators, for the entire sub-GeV to
  TeV DM mass range. This indicates that these operators need not be
  considered when analyzing data from present or future direct dark
  matter detection experiments.
\end{abstract}
	
\maketitle
	
\section{Introduction} \label{sec:intro}
	
The particle identity of Dark Matter (DM) has not yet been understood
despite overwhelming evidence for its gravitational
interactions. Dedicated direct searches exploit the possible
non-gravitational interactions of DM by looking for the recoil of target
nuclei due to elastic scattering and have already probed a large
fraction of DM mass and cross-section parameter space
\cite{Aprile:2018dbl, Amole:2019fdf, Aprile:2019dbj,
  Abdelhameed:2019hmk, Akerib:2016vxi, Akerib:2017kat, Cui:2017nnn,
  Agnese:2018gze, Agnes:2018fwg, Petricca:2017zdp}.
	
Much of the direct detection search strategy for roughly three decades
has been focused on a category of DM candidates known as Weakly
Interacting Massive Particles (WIMPs).  WIMPs with weak-scale masses,
produced in the early universe via freeze-out from the thermal plasma
\cite{Kolb:1990vq}, can easily match the measured DM relic abundance
via roughly electroweak-strength couplings to Standard Model (SM)
particles. Most analyses assume DM-nucleon interactions to be
dominated by only two operators, which describe Spin-Independent (SI)
or Spin-Dependent (SD) interactions in the limit of vanishing WIMP
velocity. The leading bounds on SI DM-nucleon cross section and SD
DM-neutron cross section are currently set by XENON1T
\cite{Aprile:2018dbl, Aprile:2019dbj}; for a $30$ GeV WIMP, the bounds
are $ 4.1 \times 10^{-47} \, \text{cm}^2 $ and
$ 6.3 \times 10^{-42} \, \text{cm}^2 $, respectively. The best direct
detection bound on the SD DM-proton cross section comes from the
PICO-60 experiment \cite{Cui:2017nnn}; for $ 25 \GeV $ DM mass cross
sections above $ 2.5 \times 10^{-41} \, \text{cm}^2 $ are excluded.
	
The standard SI and SD operators are actually the leading (zeroth
order) terms of an EFT organized in powers of small expansion
parameters such as DM velocity $ v $ and three-momentum transfer
normalized to the nucleon mass $ \left| \vec{q\,} \right|  /m_N $. Since the DM
velocity is relatively small in the solar neighborhood,
$v/c \sim \mathcal{O}(10^{-3})$, the momentum exchange is limited to
$ \left|  \vec{q\,} \right| \lsim \min[v m_\chi, v m_A] \lsim \mathcal{O}(100 \,
\text{MeV})$; here $m_\chi$ is the WIMP mass and $m_A$ is the mass of
the target nucleus. This is sizable on nuclear physics scales but
certainly well below the electroweak scale as well as the range of
WIMP masses to which these experiments are sensitive. Keeping only
the leading order terms therefore a priori seems justified.
	
Nevertheless a non-relativistic effective field theory (NREFT) was
developed which retains NLO and NNLO terms by modeling the nucleus
after taking into account the finite spatial extent of its charge and
spin densities \cite{Fan:2010gt, Fitzpatrick:2012ix, Anand:2013yka,
  Fitzpatrick:2012ib}. Working up to second order then generates a
total of 14 operators, some of which lead to novel nuclear responses
motivating the inclusion of angular-momentum dependent (LD) as well as
spin and angular-momentum dependent (LSD) interactions, along with SI
and SD. Moreover, these operators can interfere with each other and
can lead to appreciable differences from the usual exponentially
falling recoil energy spectra. It has recently been pointed out that
this affects the optimization of the experimental search strategy, in
particular by allowing events with larger recoil energy
\cite{Bozorgnia:2018jep}. Considerable effort has also been devoted to
analyzing published data in this framework \cite{Aprile:2017aas,
  Xia:2018qgs, Schneck:2015eqa, Angloher:2018fcs}.
	
The NREFT contains a total of 28 free parameters if operators
involving neutrons and protons are treated separately. This large
parameter space can be difficult to probe in its entirety. There have
been a number of global analyses of the full multi-dimensional
parameter space defining the NREFT using constraints from ongoing and
planned direct detection experiments \cite{Catena:2014uqa,
  Catena:2014epa, Kang:2018odb, Rogers:2016jrx, Liu:2017kmx,
  Catena:2015uha, Kavanagh:2015jma, Backstrom:2018plo}. One conclusion
is that direct searches constrain momentum or velocity suppressed
operators that are odd under P (parity) and T (time reversal)
transformations about as strongly as zeroth order SD interactions, but
the former have been less explored by both theorists and
experimenters. This motivates analyses of particle physics aspects
of models that can give rise to these operators in the
non-relativistic limit.
	
From a particle physics perspective, the CPT theorem implies that all
P- and T-odd NREFT operators must necessarily arise from a CP
violating theory, where C refers to charge conjugation. However, CP
violation (CPV) is highly constrained experimentally
\cite{Tanabashi:2018oca} and therefore bounds on CPV observables can
be used to constrain such NREFT operators. The neutron EDM (nEDM) is
one such observable which probes flavor diagonal CP violation and
provides one of the most stringent bounds on CPV in extensions of the
SM.  Ongoing nEDM experiments are sensitive to a possible signal which
is many orders of magnitude larger than the SM prediction
\cite{Afach:2015sja,Tanabashi:2018oca}. The current constraint is
\begin{equation} \label{eq:nEDM_limit}
| d_n | < 2.9 \times 10^{-26}  \; \text{e} \, \cdot \, \text{cm} \quad
\text{(90 \% C.L.)} \; .
\end{equation}

Various works have linked well-motivated extensions of the SM to low
energy NREFT operators \cite{DelNobile:2013sia, Gresham:2014vja,
  Catena:2017xqq, DelNobile:2018dfg, Kahlhoefer:2018knc}. In
particular, Ref.~\cite{Dent:2015zpa} lists a set of simplified models
for scalar, spinor and vector DM candidates and derives the full set
of NREFT operators in terms of the parameters for each simplified
model. In this letter, we investigate the P- and T-odd operators in
the NREFT with the help of simplified models that respect
$SU(3) \times U(1)_{\text{em}}$ invariance and are renormalizable, as
in \cite{Dent:2015zpa}. The models in this work extend the SM by a
WIMP candidate and a mediator particle. We show that in these models
the Wilson coefficients of the CPV NREFT operators and of the neutron
EDM scale with the same combination of Yukawa couplings. As a result,
we find that the constraints from the nEDM on the scattering cross
section are many orders of magnitude stronger for both CPV NREFT
operators than the current XENON1T bound and even lie below the
irreducible background from coherent neutrino scattering (``neutrino
floor'') \cite{Cabrera:1984rr, Monroe:2007xp, Billard:2013qya,
  OHare:2016pjy}. This casts doubt on the importance of these
operators for the analysis of future experiments \cite{Baudis:2012ig}
and for the formalism of NREFT itself.
	
The rest of this letter is organized as follows. In Section
\ref{sec:nreft} we introduce the NREFT formalism, the matching
procedure of a relativistic theory to NREFT operators and review the
computation of the differential event rate for elastic WIMP-nucleus
scattering. In Section \ref{sec:simplmodels} we consider two
simplified models yielding these P- and T-odd CPV NREFT operators and
compute the nEDM at 1-loop order for each case. Section
\ref{sec:results} contains our main numerical results, comparing the
nEDM and the leading direct detection bound. We conclude in Section
\ref{sec:outlook}.
	
\section{NREFT Formalism} \label{sec:nreft}
	
A non-relativistic effective field theory (NREFT) of elastic
scattering between DM and nuclei which goes beyond the strict velocity
$v \rightarrow 0$ limit \cite{Fan:2010gt, Fitzpatrick:2012ix,
  Anand:2013yka, Fitzpatrick:2012ib} offers a parametrically richer
description than the traditional relativistic four-field effective
operator approach. Since an incoming DM particle striking the target
nuclei is fairly slow in the target rest frame,
$v/c \sim \mathcal{O}(10^{-3})$, an NREFT is well suited to describe
the scattering process. We briefly review this formalism for
completeness.
	
We begin by discussing the basis of NREFT operators that describe the
interactions of a WIMP $\chi$ with a nucleon $N$ at low relative
velocities. Galilean symmetry restricts the operator basis to consist
of quantities which are invariant under Galilean boosts. The
three-momentum transfer $ \vec{q} $ and relative initial velocity
$ \vec{v} = \vec{v}_{\chi, \text{in}} - \vec{v}_{N, \text{in}} $
satisfy this criterion, as does any quantity that is computed only
from these three-vectors.  Since any two relative velocities are
sufficient to describe the scattering, it is helpful to consider
orthogonal combinations such that the scalar product of the two
invariants vanishes. The transverse velocity
$\vec{v}^{\perp} \equiv \vec{v} + \frac{\vec{q}}{2\mu_N}$ is by
construction orthogonal to the momentum transfer $\vec{q}$, where
$\mu_N = m_\chi m_N / (m_\chi + m_N)$ is the reduced mass of the
WIMP-nucleon system; $\vec{v}^{\perp} \cdot \vec{q} = 0 $ follows from
energy and momentum conservation.
	
Along with the two three-vectors $ \vec{q} $ and $ \vec{v}^{\perp} $,
the scattering can be fully described by the spins of the target
nucleon $ \vec{S}_N $ and of the DM particle $ \vec{S}_\chi $ (when
non-zero), which are also invariant under a Galilean
transformation. The operator basis constructed so far is not so
convenient for the construction of an interaction Hamiltonian since it
is not entirely Hermitean; for instance the momentum transfer
$ \vec{q} $ is anti-Hermitean. Therefore, the most general
non-relativistic interactions for elastic DM--nucleus scattering can
be written down as functions of the following four Hermitean, Galilean
invariant quantities:
\begin{equation} \label{eq:builblocks}
i\, \vec{q}, \quad \vec{v}^{\perp}, \quad \vec{S}_N, \quad \vec{S}_\rchi \, .
\end{equation}
A set of linearly independent operators $ \mathcal{O}_i $ can be
constructed using the building blocks in
Eq. \eqref{eq:builblocks}. Table \ref{table:operatorlist} lists the
set of operators which are obtained when terms up to second order in
momentum exchange $ \vec{q} $ are retained. The operators $\Oone$ and
$\Ofour$ correspond to the traditional SI and SD interaction,
respectively.
	
{ \renewcommand{\arraystretch}{2.5}
\begin{table}[ht]
\centering
\begin{tabular}{l@{\hskip 0.15in}l@{\hskip 0.15in}l}
$\Oone = 1_\rchi 1_N$; & $\mathcal{O}_6 = \left( \dfrac{\vec{q}}{m_N} \cdot \vec{S}_N \right) \left( \dfrac{\vec{q}}{m_N}
\cdot \vec{S}_\rchi \right)$; &  $\mathcal{O}_{10} = i \dfrac{\vec{q}}{m_N} \cdot \vec{S}_N$; \\
 $\mathcal{O}_3 = i \vec{S}_N \cdot \left( \dfrac{\vec{q}}{m_N} \times \vec{v}^{\perp} \right)$;
                   & $\mathcal{O}_7 = \vec{S}_N \cdot \vec{v}^{\perp} $;
                   & $\mathcal{O}_{11} = i \dfrac{\vec{q}}{m_N} \cdot \vec{S}_\rchi $; \\
  $\mathcal{O}_4 = \vec{S}_{\rchi} \cdot \vec{S}_{N}$;
                   & $\mathcal{O}_8 = \vec{S}_\rchi \cdot \vec{v}^{\perp} $;
                   & $\mathcal{O}_{12} = \vec{S}_\rchi \cdot (\vec{S}_N \times \vec{v}^{\perp})$; \\
  $\mathcal{O}_5  = i \vec{S}_\rchi \cdot  \left( \dfrac{\vec{q}}{m_N} \times \vec{v}^{\perp} \right)$;
                   & $\mathcal{O}_9 = i \vec{S}_\rchi \cdot \left(\vec{S}_N \times
                         \dfrac{\vec{q}}{m_N} \right)$;
                   & $\mathcal{O}_{13} = i ( \vec{S}_\rchi \cdot \vec{v}^{\perp} )
				\left(\dfrac{\vec{q}}{m_N} \cdot \vec{S}_N \right) $;
\end{tabular}
	$\mathcal{O}_{14} = i (\vec{S}_N \cdot \vec{v}^{\perp}) \left( \dfrac{\vec{q}}{m_N} \cdot \vec{S}_\rchi \right)$
\setlength{\tabcolsep}{15pt}
\label{OperatorList}
\caption{List of operators in the NREFT for elastic WIMP-nucleon
  scattering. We adopt the conventions of \cite{Anand:2013yka} by
  defining the operators normalized by the nucleon mass $m_N$ in order
  to have a dimensionless basis. We omit the invariant
  $\mathcal{O}_2 = v_{\perp}^2 $ because it is a second order
  correction to the SI operator $\mathcal{O}_1$, as well as
  $\mathcal{O}_{15} = - \left( \vec{S}_{\rchi} \cdot
  \frac{\vec{q}}{m_N} \right) \left( (\vec{S}_N \times \vec{v}^{\perp})\cdot
  \frac{\vec{q}}{m_N} \right)$ since it generates a cross section of order
  $ v_T^6 $, which is $ \text{N}^3 $LO.}
\label{table:operatorlist}
\end{table} }
	
In deriving the NREFT operators, no requirement to obey discrete
symmetries such as invariance under spatial parity P and time reversal T
transformations has been imposed.\footnote{Since antiparticles do not
  occur in the non-relativistic limit, C transformations do not play a
  role.} Therefore, the operators in Table \ref{table:operatorlist}
can be classified according to their P and T quantum numbers. Since
velocities and three momenta are odd under both P and T whereas spin
is odd under T but even under P, the following operators are both 
P-odd and T-odd:
\begin{equation} \label{eq:oddops}
\mathcal{O}_{10} \; , \; \mathcal{O}_{11} \; , \; \mathcal{O}_{12}
\quad \quad \text{P-odd and T-odd}\,;
\end{equation}
note that the anti-linear transformation T maps a complex number to
its complex conjugate.
	
In order to match some UV-complete or simplified model for DM onto the
NREFT, a two-step procedure is used. In the first step one integrates
out the heavy mediator field(s), leading to a still relativistic but
non-renormalizable description in terms of four-field DM-quark
operators. If WIMPs are spin-$1/2$ fermions the nucleon and DM
bilinears can be formed using a basis of 16 linear hermitean matrices
$ \Gamma_{\chi, N} \in \{I_4, i \gamma^5, \gamma^\mu, \gamma^\mu
\gamma^5, \sigma^{\mu \nu} \}$. For scalar DM the Lorentz structure
appearing in the scalar bilinear can be\footnote{In general a bilinear
  with two derivatives can also contribute; however, this does not
  happen for the models we discuss below.}
$ \{ I, i \overleftrightarrow{\partial_\mu}/m_S \} $ and the nucleon
bilinear can be
$ \{ I_4, i \gamma_5, \gamma^\mu, \gamma^\mu \gamma_5\} $. In either
case only combinations are allowed where all Lorentz indices are
contracted between the two bilinears. By convention the scalar DM
bilinears are multiplied with a factor of the scalar mass $ m_S $
i.e. $ m_S \, S^\dagger \, \Gamma_{S} \, S $ such that the Wilson
coefficients obtained have the same mass dimension as those of the
fermionic WIMPs.
	
In the non-relativistic limit these dimension-6 four field operators
composed of a nucleon bilinear and a DM bilinear reduce to
$ \mathcal{O}_N $ and $ \mathcal{O}_\rchi $ respectively; their
products can be matched on to NREFT operators
$ \mathcal{O}_{i, \text{NR}} $ of Table~\ref{table:operatorlist} with
Wilson coefficients $ c_i^N $ (in the isospin basis) of mass dimension
$ \text{GeV}^{-2} $, with
$ \mathcal{O}_{i, \text{NR}} \equiv \mathcal{O}_\rchi \cdot
\mathcal{O}_N $.
	
Any relativistic theory with a DM candidate can thus be matched on to
a particular linear combination of the NREFT operators. The rate,
differential in the recoil energy $E_R$, of elastic WIMP--nucleus
scattering events per unit time and unit detector mass, $dR/dE_R$, is
then given by
\begin{equation} \label{eq:diffrate}
\dfrac{dR}{dE_R} = N_T \dfrac{\rho_{\rchi} m_A}{2 \pi m_{\rchi}} 
\int\limits_{v_{\text{min}}}^{v_{\text{max}}} \dfrac{f(v)}{v} \; \dfrac{1}{2 j_{\rchi} + 1} \, \dfrac{1}{2 j_{A} + 1} \, 
\sum_{\text{spins}} |\mathcal{M}_{{\text{sc}}}|^2 \; d^3 v \;.
\end{equation}
Here $N_T$ is the number of target nuclei in the detector,
$\rho_{\rchi}$ is the local DM density, $m_\rchi$ is the DM mass,
$m_A$ is the mass of the target nuclei\footnote{If the detector
  contains different types of nuclei, a sum over all isotopes is
  understood in eq.(\ref{eq:diffrate}).}, $f(v)$ is the DM halo
velocity distribution in the laboratory frame, $ v_{\text{min}} $ is
the minimum velocity required to cause a nuclear recoil $ E_R $,
$ v_{\text{max}} $ is related to the galactic DM escape velocity, and
$j_{\rchi}$ and $j_{A}$ are the total spin of the DM and nucleus,
respectively. Finally, the matrix element $\mathcal{M}_{\text{sc}}$ is
given by \cite{Anand:2013yka}:
\begin{align} \label{eq:scatprob}
\dfrac{1}{2 j_{\rchi} + 1} \, \dfrac{1}{2 j_{A} + 1} \sum_{\text{spins}} |\mathcal{M}_{{\text{sc}}}|^2 =
\sum_{\substack{k' = M,\Sigma^{''},\\\Sigma^{''}}} R_k^{NN'}(v^2, \vec{q}^{\, 2}) \,
W_k^{NN'}(\vec{q}^{\, 2} b^2) + \sum_{\substack{k' = \Delta,\Delta \Sigma^{'}, \\ \Phi^{''},\Phi^{''}M}}
\dfrac{\vec{q}^2}{m_N^2} R_k^{NN'}(v^2, \vec{q}^{\, 2}) \,
W_k^{NN'}(\vec{q}^{\, 2} b^2) \;.
\end{align}
The result (in the isospin basis) has been factorized into WIMP
response functions $R_k$ and nuclear response functions $W_k$. Both
response functions depend on the momentum transfer. The former also
depend on the relative WIMP-nucleon velocity and encode the particle
physics of the scattering process, and thus explicitly depend on the
product of Wilson coefficients $c^N_i c^{N'}_j$ and can be computed
perturbatively; see Ref.~\cite{Anand:2013yka} for a list of explicit
expressions. The latter encode the target-dependent nuclear physics;
they are functions of $y \equiv (qb/2)^2$, where $b$ is the nuclear
size. The sum over $k$ in Eq. \eqref{eq:scatprob} involves the
enlarged set of five independent and two interfering responses: the
standard SI $M$, the transverse SD $\Sigma'$, the longitudinal SD
$\Sigma^{''}$, the angular momentum dependent (LD) $\Delta$ and the
angular momentum and spin dependent (LSD) $\Phi^{''}$ whereas
interference of the angular momentum and spin dependent response with
the standard SI leads to $\Phi^{''}M$ and the transverse SD and the
angular momentum response interfere to give rise to
$\Delta \Sigma^{'}$. Five of the seven responses are accompanied by a
factor of $\vec{q}^{\; 2}/m_N^2 $, a parameter related to the relative
velocities of nucleons bound in the nucleus, further reflecting that
the NREFT operators associated with the responses are all suppressed
by at least $\vec{q}^{\, 2}/m_N^2$.
	
If the Wilson coefficients in the final non-relativistic effective
action are all of the same order of magnitude, the NREFT operators
listed in Table \ref{table:operatorlist} are not equally relevant for
scattering. The traditional SI operator $\Oone$ then dominates over
the rest since it is enhanced by a factor of $ A^2 $, where $A$ is the
nucleon number, and has no velocity suppression. Naively, one would
conclude that the other remaining momentum/velocity independent
operator $\Ofour$ would be the dominant contribution to scattering in
the absence of $\Oone$; however, that is not always the case. Recall
that the momentum transfer involved in elastic DM-nucleon scattering
can be of the order a hundred MeV, $ |\vec{q}| \sim 100 \text{ MeV}
$. Once the target nucleus can be resolved into its constituent
nucleons, there will be contributions which are only suppressed by the
normalized momentum transfer $\vec{q}^{\,2}/m_N^2$, which can be
$\mathcal{O}(10^{-2})$, rather than by the squared velocity
$\vec{v}_T^{\,2} \sim \mathcal{O}(10^{-6})$. Therefore the contribution
from momentum transfer suppressed NREFT operators frequently dominates
over those from velocity suppressed ones, at least for relatively
large recoil energies. As a result, in the absence of $ \Oone $ the SI
P- and T-odd operator $\Oeleven$ often dominates scattering instead of
$\Ofour$, {\em if} the corresponding Wilson coefficients are of
similar size. This explains why numerical scans \cite{Catena:2014uqa,
  Catena:2014epa} found fairly stringent constraints on the Wilson
coefficient of these operators.

\section{DM Simplified Models} \label{sec:simplmodels}
		
Simplified models of DM provide a minimal framework to explore its
phenomenology across direct, indirect and LHC searches. In these
models the SM is augmented with a WIMP candidate and a mediator via
which quarks and/or leptons can interact with WIMPs. Although
renormalizable and invariant under $SU(3)_{C} \times U(1)_{\text{em}}$
gauge transformations, they usually ignore the $SU(2)$ part of the SM
gauge group; they are therefore not UV complete theories. However,
since simplified models contain a relatively small number of
parameters they can be powerful tools to interpret experimental
results and explore the complementarity of different searches.
	
Ref. \cite{Dent:2015zpa} has constructed a set of simplified models by
exhaustively listing different WIMP and mediator spins and matched the
resulting relativistic Lagrangians to the set of non-relativistic
operators listed in Table \ref{table:operatorlist}. In these models
the WIMP is an $SU(3)_C \times U(1)_{\text{em}}$ singlet; the
interactions are further constrained by the requirement that the WIMP
cannot decay.
	
Here we focus on two simplified models with color-triplet mediators
which in the NR limit give rise to the leading order P- and T-odd
operators $\Oeleven,\, \Oten$ and $\mathcal{O}_{12}$; $\Oeleven$ is
spin-independent while $\Oten$ and $\mathcal{O}_{12}$ are
spin-dependent. In the following subsections we describe these models
and use the experimental upper bound on the neutron EDM to constrain
their direct detection prospects.
	
\subsection{Model I}
	
Model I contains a complex spin-$0$ WIMP $S$ and one or more heavy quark-like
mediator(s) $Q$, both of which are odd under a discrete symmetry
$\mathbb{Z}_2$ to forbid dark matter decay. The SM quarks $q$ can have two
new Yukawa-type interactions with the WIMP and the mediator(s), via
scalar couplings $y_1^q$ and pseudoscalar couplings $y_2^q$, where the
superscript denotes the quark flavor index $ q $. The Lagrangian thus
reads:
\begin{align} \label{eq:Lag_Model_I}
\mathcal{L}^{\text{Model I}} = & \, \mathcal{L}_{SM} + \, \partial_\mu S^\dagger
\partial_\mu S - m_S^2 S^\dagger S - \lambda_S (S^\dagger S)^2 \nonumber \\
&+ i\bar{Q_k} \slashed{D} Q_k - m_{Q_k} \bar{Q_k} Q_k \nonumber \\
&-S \bar{Q_k} (y_1^q + y_2^q \gamma^5) q_l \: - \: S^{\dagger} \bar{q_l}
(y_1^{q \, \dagger} - y_2^{q \, \dagger} \gamma^5) Q_k \; .
\end{align}
$U(1)_{\rm em}$ invariance implies that at least two mediators are required
if the WIMP $S$ is to couple to both up- and down-type quarks. Moreover,
a mediator $Q_k$ with nonvanishing couplings to quarks of different
generations will give rise to flavor changing neutral current (FCNC)
processes at one-loop order. For example, a $Q_k$ coupling to both
$d$ and $s$ quarks will contribute to $K^0 - \overline{K^0}$ mixing
via box diagrams with $Q_k$ and $S$ in the loop. We will therefore
assume that each mediator $Q_q$ couples only to a single SM quark
$q$ in the mass basis, so that the Yukawa coupling matrices appearing
in eq.(\ref{eq:Lag_Model_I}) reduce to diagonal matrices.

The matrix element for $s$-channel scattering
$S(p_1) + \;q(p_2) \longrightarrow S(k_1) + \; q (k_2) $ is then given
by:
%
\begin{align} 	\label{eq:Matrix_Element_Model_I}
\mathcal{M}_{Sq \rightarrow Sq} &= \dfrac{m_{Q_q}}{m_{Q_q}^2 - m_S^2}
\Big[ (\abs{y_1^q}^2 - \abs{y_2^q}^2) \; \bar{u}(k_2) \, u(p_2)  -
2 \, \text{Im}(y_1^q y_2^{q \, \dagger}) \; \bar{u}(k_2) \, i\gamma^5 \, u(p_2)\Big]
\nonumber \\
&+ \dfrac{1}{m_{Q_q}^2 - m_S^2} \, (\abs{y_1^q}^2 + \abs{y_2^q}^2 )
\Big[  \,m_q \, \bar{u}(k_2) \, u(p_2) + \; \bar{u}(k_2) \, \dfrac{\slashed{p_1}
+\slashed{k_1}}{2}\,u(p_2) \Big]\nonumber \\
&+ \dfrac{1}{m_{Q_q}^2 - m_S^2} \; 2 \, \text{Re}(y_1^q y_2^{q \, \dagger})\;
\Big[ \, \bar{u}(k_2) \, \dfrac{\slashed{p_1} +\slashed{k_1}}{2} \gamma^5 \,
u(p_2) \Big] \; .
\end{align}
%
Here we have neglected the mass of the incoming quark as well as terms
of order $|\vec{q}|^2$ in the denominator of the $Q_q$ propagator, but
we have kept these terms in the numerator; note that at tree-level
only scattering on $u,\, d$ and $s$ quarks contribute to WIMP-nucleon
scattering. Moreover, we have used the Dirac equation and $4$-momentum
conservation to write the matrix element in a form that is symmetric
in WIMP momenta. This facilitates the construction of the
corresponding relativistic effective Lagrangian, which results when
the mediator $Q_q$ is integrated out. Defining the Hermitean derivative
on the complex scalars as
$i S^\dagger \overleftrightarrow{\partial_\mu} S \equiv
\frac{i}{2}(S^\dagger \partial_\mu S - S \partial_\mu S^\dagger)$, we
have:
%
\begin{equation} \label{eq:Effective_Operators_Model_I}	
\mathcal{L}_{\text{eff}}^{\text{Model I}} \supset c^{q,d5}_1 \;
(S^\dagger S) \, \bar{q} \, q \; + \; c^{q,d5}_{10} \;
(S^\dagger S) \, \bar{q} \, i \gamma^5 \, q \; + \;
c_1^{q,d6} \; (i S^\dagger \overleftrightarrow{\partial_\mu} S) \, \bar{q} \,
\gamma^\mu \, q \; + \; c_7^{q,d6} \; (i S^\dagger
\overleftrightarrow{\partial_\mu} S) \, \bar{q} \, \gamma^\mu \gamma^5 \, q\,.
\end{equation}
%
The subscripts on the Wilson coefficients denote the NREFT operator
that the corresponding relativistic effective operator reduces to and
the superscripts denote the mass dimension of the corresponding
relativistic effective operator.
	
{\renewcommand{\arraystretch}{2.5}
\begin{table}[h!]
\begin{tabular}{c c c}
$ S^\dagger \Gamma_{S} S  \; \bar{N} \, \Gamma_{N} \,N $ & & $ c^q_i \; \mathcal{O}_i $ \\ \hline
$ c^{q,d5}_1 S^\dagger S \; \bar{q} q$  & $\longrightarrow$ & $\left(  \dfrac{m_{Q_q}}{m_S}
	\dfrac{\abs{y_1^q}^2 - \abs{y_2^q}^2}{m_{Q_q}^2 - m_S^2}+ \dfrac{m_q}{m_S}
	\dfrac{\abs{y_1^q}^2 + \abs{y_2^q}^2}{m_{Q_q}^2 - m_S^2} \right)
	f_{Tq}^N \, \mathcal{O}_1 $ \\
$c^{q,d5}_{10} S^\dagger S \; \bar{q} i \gamma^5 q$ & $ \longrightarrow $ &
	$ \dfrac{m_{Q_q}}{m_S} \dfrac{\text{Im}(y_1^q y_2^{q \, \dagger})}{m_{Q_q}^2 - m_S^2} \,
	2\tilde{\Delta}^N \, \mathcal{O}_{10} $ \\ 
$c_1^{q,d6}  i \left(S^\dagger \overleftrightarrow{\partial_\mu} S \right) \;
	\bar{q} \gamma^\mu q $ & $\longrightarrow  $ & $ \dfrac{\abs{y_1^q}^2
	+ \abs{y_2^q}^2}{m_{Q_q}^2 - m_S^2} \, \mathcal{N}_{q}^N \, \mathcal{O}_{1} $\\ 
$c_7^{q,d6} i \left(S^\dagger \overleftrightarrow{\partial_\mu} S \right) \;
	\bar{q} \gamma^\mu \gamma^5 q  $ & $\longrightarrow $
	& $- \dfrac{\text{Re}(y_1^q y_2^{q \, \dagger})}{m_{Q_q}^2 - m_S^2} \, 2\Delta^N_q \,
	\mathcal{O}_{7}    $ \\ 
\end{tabular}
\caption{Non-relativistic reduction of effective operators in Model
  I. $ f_{Tq}^N $, $ \mathcal{N}_{q}^N $, $ \Delta^N_q $ and
  $ \tilde{\Delta}^N $ are coefficients arising due to promoting quark
  bilinear to nucleon bilinears \cite{Agrawal:2010fh,
    Dienes:2013xya}. We use the values as given in the Appendix of
  Ref. \cite{Dent:2015zpa}.}
\label{table:coeff_model_i}
\end{table}
}
	
Table \ref{table:coeff_model_i} contains the matching of the
relativistic effective operators to the NREFT operators in terms of
the parameters of Model I.  The Wilson coefficients for the
dimension-5 operators $ S^\dagger S \, \bar{q} q$ and
$ S^\dagger S \; \bar{q} i \gamma^5 q$ have been divided by a factor
of $ m_S $ such that the expressions for all DM-nucleon cross sections
contain the same factor $ \frac{\mu_{\chi N}^2}{\pi} \, (c_{i}^N)^2 $
irrespective of the mass dimension of the relativistic operator
involved. The operators $ (S^\dagger S) \; (\bar{q} q) $ and
$ i \left(S^\dagger \overleftrightarrow{\partial_\mu} S \right) \;
\bar{q} \gamma^\mu \gamma^5 q $ both reduce to the leading SI $\Oone$
operator but with different coupling combinations. Ignoring the
sub-leading $\mathcal{O}(\frac{m_q}{m_{Q_q}^2 m_S})$ contributions,
the contribution $\propto (S^\dagger S) \; (\bar{q} q) $ scales as the
difference of the absolute value squared of the Yukawa couplings
$ \abs{y_1^q}^2 - \abs{y_2^q}^2 $ whereas that
$\propto i \left(S^\dagger \overleftrightarrow{\partial_\mu} S \right)
\; \bar{q} \gamma^\mu \gamma^5 q $ scales as the sum of the absolute
value squared of the Yukawa couplings
$ \abs{y_1^q}^2 + \abs{y_2^q}^2 $; the latter contribution is
suppressed by a relative factor $m_S/m_{Q_q}$. The scalar-pseudoscalar
operator $ (S^\dagger S) \; (\bar{q} i \gamma^5 q )$ reduces to the
$ \vec{q} $ suppressed P- and T-odd SD operator $\Oten$ and
$ i \left(S^\dagger \overleftrightarrow{\partial_\mu} S \right) \;
\bar{q} \gamma^\mu \gamma^5 q $ reduces to the $ \vec{v}_\perp $
suppressed P-even, T-odd SD operator $ \mathcal{O}_7 $.
	
$ \Oone $ is the leading order operator, because it does not suffer
from any velocity suppression. It therefore contributes dominantly
when compared to $ \mathcal{O}_7 $ and $ \Oten $, {\em unless} its
Wilson coefficient is suppressed by a factor of $ 10^{-3} $ or
less. It should be emphasized that this suppression should occur for
scattering on both neutrons and protons.\footnote{For a detector
  containing a single isotope one would only need to suppress the
  coefficient $c_1$ for the specific linear combination of neutrons
  and protons determined by the target. However, the Xenon detectors,
  which currently give the best bounds, contain several isotopes, and
  unsuppressed $\mathcal{O}_1$ constraints from Germanium are still
  significantly stronger than $\mathcal{O}_{10}$ constraints from
  Xenon.} Since the relevant hadronic matrix elements are different
for neutrons and protons, one {\em cannot} arrange both cancellations
with only a single mediator. In other words, the P- and T-odd
operators can only be significant in this model if one introduces at
least two mediators. Suppressing all contributions from $\Oone$ would
then require two relations between the twelve couplings $y_1^q, y_2^q$
and six masses $m_{Q_q}$ to hold to better than one part in $10^3$.

On the other hand, if one introduces mediators with charge $1/3$ and
with charge $2/3$, one can require the Lagrangian
(\ref{eq:Lag_Model_I}) to respect strong isospin invariance. In this
case the cancellations for protons and neutrons are in fact (almost)
the same. In the most symmetric set-up where each SM quark $q$ is
assigned a mediator $Q_q$, the couplings $y_1^q \equiv y_1$ and
$y_2^q \equiv y_2$ are flavor-universal, and all mediators have the
common mass $m_Q$ the contributions from $\Oone$ vanish if
\begin{align} \label{eq:forbid_O1_Model_I}
\abs{y_1}^2 = \left(  \dfrac{1 - \dfrac{\mathcal{N}^N}{f^N_T}
\, \dfrac{m_S}{m_Q}} {1 + \dfrac{\mathcal{N}^N}{f^N_T} \, \dfrac{m_S}{m_Q}}
\right) \abs{y_2}^2\,.
\end{align}
Here $ f^N_T \equiv \sum_{q} \bra{\bar{N}} \bar{q} q \ket{N}$ is the
contribution of light and heavy quarks to the nucleon mass (scalar
nucleon bilinear) and
$ \mathcal{N}^N \equiv \sum_{q} \bra{\bar{N}} \bar{q} \gamma^\mu q
\ket{N} $ is the number of valence quarks in the nucleon (vector
nucleon bilinear).  For our assumption of flavor-universal couplings,
$ f^N_T $ is given by
\begin{align} \label{eq:fnT}
f^N_T = \sum\limits_{q=u,d,s} \frac{m_N}{m_q} f^N_{Tq} + 
\frac{2}{27} \left( 1 - \sum\limits_{q'=u,d,s} f^N_{Tq'} \right) 
\sum\limits_{q=c,b,t} \frac{m_N}{m_q} \; ,
\end{align}
where the heavy quark contribution is due to the trace anomaly of the
energy momentum tensor \cite{Shifman:1978zn}.\footnote{The heavy quark
  contributions to $ f^N_T $ in Eq.\eqref{eq:fnT} assumes
  $m_Q \gg m_q$ so that the mediators can be integrated out
  consistently, leaving only SM quarks behind. This may not be a good
  approximation for the top quark and its mediator. For
  flavor-universal couplings the contributions from heavy SM quarks is
  in any case negligible. However, one might also entertain the
  possibility that the $|y_i^q|^2$ scale $\propto m_q$, so that the
  $Q-S$ two-point function corrections to the SM quark masses scale
  like $m_q$; in this case the contributions from heavy SM quarks
  would be comparable to that in models where the WIMP interacts with
  nucleons via Higgs exchange.} The vector nucleon bilinear
coefficient $ \mathcal{N}^N $ is $3$ for both neutrons and protons
reflecting the number of valence quarks. Using the values for
$f_{Tq}^N $ given in Ref. \cite{Ellis:2018dmb}, the ratio of the
nucleon bilinears appearing in Eq.\eqref{eq:forbid_O1_Model_I} is
\begin{align}
\dfrac{\mathcal{N}^N}{f^N_T} = \begin{cases} 0.212^{+0.043}_{-0.038} ,& 
N= n \\ 0.219^{+0.051}_{-0.044} ,& N=p \end{cases}\; .
\end{align}
Note that the stability of the WIMP $S$ requires $m_S \leq m_Q$, hence
eq.\eqref{eq:forbid_O1_Model_I} relates the two Yukawa couplings via
an $ \mathcal{O}(1) $ factor. From the simplified model point of view
there is a priori no reason why Eq.\eqref{eq:forbid_O1_Model_I} (or a
suitable modification thereof) should hold, i.e. in almost all of the
parameter space the contribution from the traditional operator $\Oone$
will in fact dominate.\footnote{The contribution to the Wilson
  coefficient of $\Oone$ which is suppressed by $m_S/m_Q$ has not been
  included in \cite{Dent:2015zpa}. If this contribution is neglected,
  the Wilson coefficient vanishes for $ |y_1| = |y_2|$, which could be
  motivated via chiral symmetry. However, an otherwise unsuppressed
  contribution from $(m_S/m_Q) \Oone$ would still dominate over
  contributions from $\mathcal{O}_7$ and $\Oten$ unless
  $m_Q \gsim 10^3 m_S$ in which case the scattering cross section is
  anyway very small.} 
	
Imposing Eq.\eqref{eq:forbid_O1_Model_I}, $ \Oten $ remains as
dominant operator since its contribution to the scattering matrix
element is only suppressed by $ \frac{\vec{q}}{m_n} $, as compared to
a suppression $ \vec{v}_{\perp} $ in $ \mathcal{O}_7 $, as discussed
in Section \ref{sec:nreft}.

\begin{figure*}[t!]
\centering
\begin{subfigure}[b]{0.45\textwidth}
\centering
\includegraphics[width=0.6\linewidth]{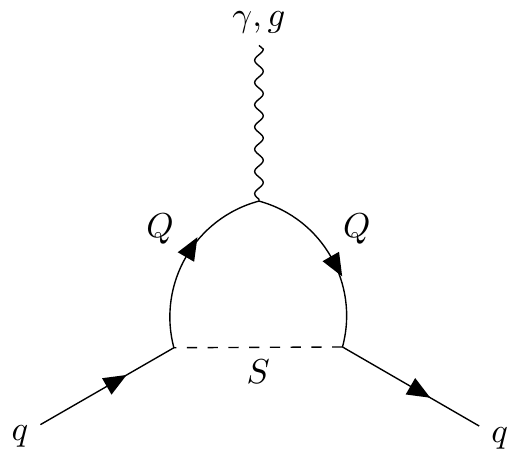}
	\caption{Model I}
	\label{fig:EDM_Model_I}
\end{subfigure}%
\begin{subfigure}[b]{0.45\textwidth}
	\centering
	\includegraphics[width=0.6\linewidth]{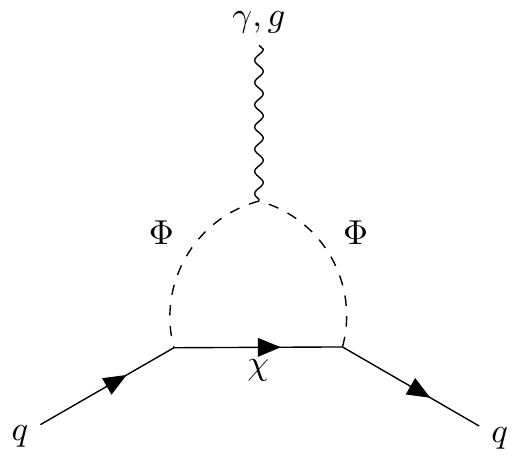}
	\caption{Model II}
	\label{fig:EDM_Model_II}
\end{subfigure}
\caption{Feynman diagrams for quark EDMs and color-EDMs in Model I
  (left) and II (right).  These diagrams have been drawn using
  TikZ-Feynman \cite{Ellis:2016jkw}.}
\label{fig:EDM_1loop}
\end{figure*}
	
The quark EDM, $ d_q $, can be calculated as the coefficient of a
dimension-$5$ P- and T-odd interaction term
$ (-i/2) \, \bar{q} \sigma_{\mu \nu} \gamma_5 q \, F^{\mu \nu} $ at
vanishing momentum transfer.  Since the mediator $ Q $ is charged
under $SU(3)_C$, it can couple to gluons, hence non-vanishing
chromo-quark EDMs might also be generated. These are calculated
similar to quark EDMs, by finding the coefficient of
$ (-i/2) \, \bar{q} \sigma_{\mu \nu} t_a \gamma_5 q \, G^{\mu \nu}_a $
at vanishing momentum transfer. From the left diagram in
Fig.~\ref{fig:EDM_1loop} we calculate the quark EDM to be
\begin{align} 	\label{eq:qEDM_Model_I}
d_q |_{\text{Model I}} = \dfrac{1}{(4 \pi)^2} \; eQ_q \, m_{Q_q}\;
\text{Im} (y_1^q y_2^{q\dagger}) \; F (m_q^2, m_S^2, m_{Q_q}^2) \; ,
\end{align}
where $Q_q$ is the electric charge of the quark $q$, and hence of the
fermionic mediator (in units of the proton charge), and the loop
function $ F (m_q^2, m_S^2, m_Q^2) $ is given by
\begin{equation} \label{eq:Loop_Function_Model_I}
F(m_q^2, m_S^2, m_Q^2) = \int_{0}^{1} dz \dfrac{(1-z)^2}
{z^2 m_q^2 + z(m_S^2 - m_Q^2 - m_q^2) + m_Q^2} \; .
\end{equation}
The chromo-EDM, $ \tilde{d}_q $ can be obtained by replacing the
external photon with a gluon as in Fig. \ref{fig:EDM_Model_I}. The
chromo-quark EDM is thus also given by Eq. \eqref{eq:qEDM_Model_I}
after replacing $ eQ_q $ with the strong coupling $ g_s $. In the
limit $m_q \rightarrow 0$ the loop function
(\ref{eq:Loop_Function_Model_I}) varies between $1/(2m_Q^2)$ for
$m_S^2 \ll m_Q^2$ and $1/(3 m_Q^2)$ for $m_S^2 = m_Q^2$.

\subsection{Model II}

Model II introduces a fermionic WIMP $ \chi $ and one or more complex spin-0
mediator(s) $ \Phi $, both odd under a $ \mathbb{Z}_2 $ symmetry which
stabilizes $\chi$. This model is similar to Model I, but the spins of the WIMP
and the mediator have been exchanged, i.e. Model II resembles a supersymmetric
model where $\chi$ is the lightest neutralino and the mediators are squarks.
The renormalizable $SU(3)_C \times U(1)_{\rm em}$ invariant Lagrangian
is given by
\begin{align}
  \mathcal{L}^{\text{Model II}} &=  \, \mathcal{L}_{SM} + \, i\bar{\chi} \slashed{D} \chi -
                  m_\chi \bar{\chi} \chi  \nonumber \\
                  &+ (\partial_\mu \Phi_q^\dagger)(\partial^\mu \Phi_q)
                 - m_{\Phi_q}^2 \Phi_q^\dagger \Phi_q
        - \dfrac{\lambda_{\Phi_q}}{2} (\Phi_q^\dagger \Phi_q)^2 \nonumber \\
&-(l_1^q \Phi_q^\dagger \bar{\chi} q + l_2^q \Phi^\dagger_q \bar{\chi} \gamma_5 q  + h.c. )\,.
\end{align}
We have again assumed the two new types of Yukawa couplings, $l_1$ and $l_2$, to be
flavor-diagonal in order to avoid new one-loop contributions to FCNC processes
due to mediator-WIMP loops. Of course, the electric charge of the mediator $\Phi_q$
must again be the same as that of the corresponding quark $q$. Moreover, complex
phases in the new Yukawa couplings will again lead to CP violation.

The matrix element for WIMP-quark scattering via $\Phi_q$ exchange in the $s$-channel
is now given by
\begin{align}	\label{eq:Matrix_Element_Model_II}
  \mathcal{M}_{\chi q \rightarrow \chi q} &= \dfrac{1}{m^2_{\Phi_q} - m_\chi^2} \Bigl(
   l_1^q l_1^{q \, \dagger} \; [\bar{u}(k_2) v(k_1)] \, [\bar{v}(p_1) v(p_2)]
  - l_1^q l_2^{q \, \dagger} \; [\bar{u}(k_2) \gamma^5 v(k_1)] \, [\bar{v}(p_1) v(p_2)]
  \nonumber \\ 
         &+ l_1^{q \, \dagger} l_2^q \; [\bar{u}(k_2) v(k_1)] \, [\bar{v}(p_1) \gamma^5 v(p_2)]
         - l_2^q l_2^{q \, \dagger} \; [\bar{u}(k_2) \gamma^5 v(k_1)] \, [\bar{v}(p_1)
        \gamma^5 v(p_2)] \Bigr)\,.
\end{align}
%
Applying Fierz rearrangement identities from Appendix
\ref{app:Fierz_Identities} to the spinor quadrilinears appearing in
eq.(\ref{eq:Matrix_Element_Model_II}), a number of
relativistic effective operators are generated. They are listed in Table
\ref{tab:Model_II_Op_Coeff}, together with their matching to the NREFT operators in terms
of the parameters of this model.

\begin{table}[h!]
	\renewcommand{\arraystretch}{2.5}
	\begin{tabular}{l c r}
		$ \bar{\rchi} \, \Gamma_{\rchi} \, \rchi \; \bar{N} \, \Gamma_{N} \,N $ & &
		$c_i^q \; \mathcal{O}_i$ \\ \hline
		$\bar{\chi} \chi \; \bar{q} q $ & : &
		$ \dfrac{1}{4} \dfrac{\abs{l_2^q}^2 - \abs{l_1^q}^2}{m_{\Phi_q}^2 - m_\chi^2}
		f^N_{Tq} \, \mathcal{O}_1 $ \\ 
		$\bar{\chi} \chi  \; \bar{q} i \gamma^5 q $ & : & $ -\dfrac{1}{2}
		\dfrac{ \text{Im}(l_1^q l_2^{q \, \dagger}) }{m_{\Phi_q}^2 - m_\chi^2}
		\Delta \tilde{q}^N \, \mathcal{O}_{10} $\\ 
		$\bar{\chi} i \gamma^5 \chi  \; \bar{q} q $ & : &
		$ -\dfrac{1}{2} \dfrac{ \text{Im}(l_1^q l_2^{q \, \dagger}) }{m_{\Phi_q}^2 - m_\chi^2}
		\dfrac{m_N}{m_\chi} f^N_{Tq} \,\mathcal{O}_{11} $\\ 
		$\bar{\chi} i \gamma^5 \chi  \; \bar{q} i \gamma^5 q $ & : &
		$ \dfrac{1}{4} \dfrac{\abs{l_2^q}^2 - \abs{l_1^q}^2}{m_{\Phi_q}^2 - m_\chi^2}
		\dfrac{m_N}{m_\chi} \, \Delta \tilde{q}^N \mathcal{O}_{6} $\\ 
		$\bar{\chi} \gamma^\mu \chi  \; \bar{q} \gamma_\mu q $ & :  &
		$ -\dfrac{1}{4} \dfrac{\abs{l_2^q}^2 + \abs{l_1^q}^2}{m_{\Phi_q}^2 - m_\chi^2}
		\mathcal{N}^N_{q} \, \mathcal{O}_1  $\\ 
		$\bar{\chi} \gamma^\mu \gamma^5 \chi  \; \bar{q} \gamma_\mu q $ & : &
		$ \dfrac{\text{Re}(l_1^q l_2^{q \, \dagger})}{m_{\Phi_q}^2 - m_\chi^2} \mathcal{N}_q^N \,
		(\mathcal{O}_8 + \mathcal{O}_9) $ \\ 
		$ \bar{\chi} \gamma^\mu \chi  \; \bar{q} \gamma_\mu \gamma^5 q $ & : & $
		\dfrac{\text{Re}(l_1^q l_2^{q \, \dagger})}{m_{\Phi_q}^2 - m_\chi^2} \Delta_q^N \,
		(-\mathcal{O}_{7} + \frac{m_N}{m_\chi}\mathcal{O}_{9}) $\\ 
		$ \bar{\chi} \gamma^\mu \gamma^5 \chi \; \bar{q} \gamma_\mu \gamma^5 q $ & : &
		$ -  \dfrac{\abs{l_2^q}^2 + \abs{l_1^q}^2}{m_{\Phi_q}^2 - m_\chi^2}
		\Delta_q^N \mathcal{O}_4 $ \\ 
		$ \bar{\chi} \sigma^{\mu \nu}  \chi \; \bar{q} \bar{\chi} \sigma_{\mu \nu}  \chi q $ & : &
		$ \dfrac{\abs{l_2^q}^2 - \abs{l_1^q}^2}{m_{\Phi_q}^2 - m_\chi^2}
		\delta_q^N \mathcal{O}_4 $\\
		$ \bar{\chi} \sigma^{\mu \nu}  \gamma^5 \chi \; \bar{q} \bar{\chi} \sigma_{\mu \nu}  \chi q $ &
		:  & $\dfrac{2 \text{Im}(l_1^q l_2^{q \, \dagger})}{m_{\Phi_q}^2 - m_\chi^2}
		\delta^N_q \left( \mathcal{O}_{11} -  \frac{m_N}{m_\chi}\mathcal{O}_{10}
		- 4 \mathcal{O}_{12} \right) $ 
	\end{tabular}
\caption{Reduction to effective non-relativistic operators in Model
  II. The left column gives the relativistic 4-fermion operators that
  appear in the matrix element for WIMP-nucleon scattering, and the
  right column gives the corresponding NREFT operators including the
  coefficient appearing in this scattering matrix element.
  $ f_{Tq}^N $, $ \Delta \tilde{q}^N $, $ \mathcal{N}^N_{q} $,
  $ \Delta_q^N $ and $ \delta_q^N $ are coefficients arising due to
  promoting quark bilinear to nucleon bilinears \cite{Agrawal:2010fh,
    Dienes:2013xya}. We use the values as given in the Appendix of
  Ref. \cite{Dent:2015zpa}.}
\label{tab:Model_II_Op_Coeff}
\end{table}	

We see that Model II generates the following NREFT operators:
$ \Oone $, $ \Ofour$, $ \mathcal{O}_6 $, $ \mathcal{O}_7 $,
$ \mathcal{O}_8 $, $ \mathcal{O}_9 $, $ \Oten $, $ \Oeleven $ and
$ \mathcal{O}_{12} $. Referring back to Table~\ref{table:operatorlist},
$ \mathcal{O}_6, \, \mathcal{O}_7, \mathcal{O}_9, \mathcal{O}_{10} $
and $ \mathcal{O}_{12} $ are all spin-dependent but are suppressed by
at least one factor of $ \vec{v}^\perp \sim 10^{-3}$ or
$\vec{q}/m_N$. They can therefore be neglected relative to
$\mathcal{O}_4$, which did not contribute in Model~I. Moreover,
$\mathcal{O}_8$ is suppressed by $\vec{v}^\perp \sim 10^{-3}$ and can
thus also be neglected.  As in Model~I the dominant term will come
from $\Oone$ {\em unless} its coefficient is tuned to be tiny. The
contributions from $\Ofour$ and $\Oeleven$ might be roughly
comparable: the latter is suppressed by a single power of
$\vec{q}/m_N$, but enhanced by the coherence factor $A$ since it does
not depend on the spin of the nucleus.

As in Model~I, there are two contributions that reduce to $\Oone$: the
scalar-scalar bilinear $ \bar{\chi} \chi \; \bar{q} q $ contributes
proportional to the difference of the absolute squares of the Yukawa
couplings $ \abs{l_2^q}^2 - \abs{l_1^q}^2 $ whereas the vector-vector
bilinear $ \bar{\chi} \gamma^\mu \chi \; \bar{q} \gamma_\mu q $
contributes $ \propto \left(\abs{l_2^q}^2 + \abs{l_1^q}^2\right) $. The
total coefficient of $ \Oone $ can therefore again be made to vanish
by an explicit cancellation. Assuming universal couplings and masses
for all mediators, the cancellation condition is
\begin{align} \label{eq:forbid_O1_Model_II}
  \abs{l_1}^2 = \left(  \dfrac{1 -  \frac{\mathcal{N}^N}{f^N_T} }
  {1 + \frac{\mathcal{N}^N}{f^N_T} } \right) \abs{l_2}^2 \; .
\end{align}
This relation depends on the same ratio of nuclear matrix elements as
the analogous relation (\ref{eq:forbid_O1_Model_I}) for Model~I.
We note that these universality assumptions forbid to also tune the
coefficient of $\Ofour$ to zero. This could be arranged for general
couplings; however, if the new interactions violate strong isospin,
the coefficients of $\mathcal{O}_1^p$ and $\mathcal{O}_1^n$ would have
be tuned to zero separately.

We see from Table~\ref{tab:Model_II_Op_Coeff} that the coefficient of
the leading new operator $\Oeleven$ is proportional to the relative
phase between the two new Yukawa couplings; this is true also for the
coefficients of the other two P- and T-odd operators, $\Oten$ and
$\mathcal{O}_{12}$.  Since this phase signals CP violation, we expect
it to show up in the the quark EDM and CEDMs. The relevant diagrams in
this setup are depicted in Fig.~\ref{fig:EDM_Model_II}. The resulting
quark EDM is given by
\begin{equation} \label{eq:qEDM_Model_II}
  d_q|_{\text{Model II}} = \dfrac{1}{(4 \pi)^2} \;e Q_q \;  m_\chi \;
  \text{Im}(l_1 l_2^\dagger) \; G (m_q^2, m_\Phi^2, m_\chi^2) \,.
\end{equation}
Here $ Q_q $ is again the electric charge of $q$, which is equal to the charge
of $\Phi_q$, and the loop function $ G (m_q^2, m_\Phi^2, m_\chi^2) $ is given by
\begin{equation} \label{eq:Loop_Function_Model_II}
  G (m_q^2, m_\Phi^2, m_\chi^2) = \int_{0}^{1} dz \: \dfrac{z(1-z)}
  { z^2 m_q^2 + z(m_\chi^2 - m_\Phi^2 - m_q^2) + m_\Phi^2 } \; .
\end{equation}
Again, the quark CEDM can be obtained by replacing $ e Q_q $ by the
strong coupling $ g_s $ in Eq.~\eqref{eq:qEDM_Model_II}. The EDMs in
Eq. \eqref{eq:qEDM_Model_I} and \eqref{eq:qEDM_Model_II} are
proportional to the mass of the fermion running in the loop, since the
dipole operators violate chirality. Moreover, we see the same
CP-violating combination of couplings as in the coefficient of the P-
and T-odd NREFT operators. The loop function $G$ varies from
$1/(2 m_\Phi^2)$ for $m_\chi^2 \ll m_\phi^2$ to $1/(6 m_\Phi^2)$ for
$m_\chi = m_\Phi$.  We note that our loop calculations in
\eqref{eq:qEDM_Model_I} and \eqref{eq:qEDM_Model_II} are in complete
agreement with earlier results in the literature, for instance the
model independent calculation of EDMs of ref.~ \cite{Abe:2017sam}.
	
\section{Results and Discussion} \label{sec:results}
	
\begin{figure*}
\centering
\includegraphics[width=0.45\linewidth]{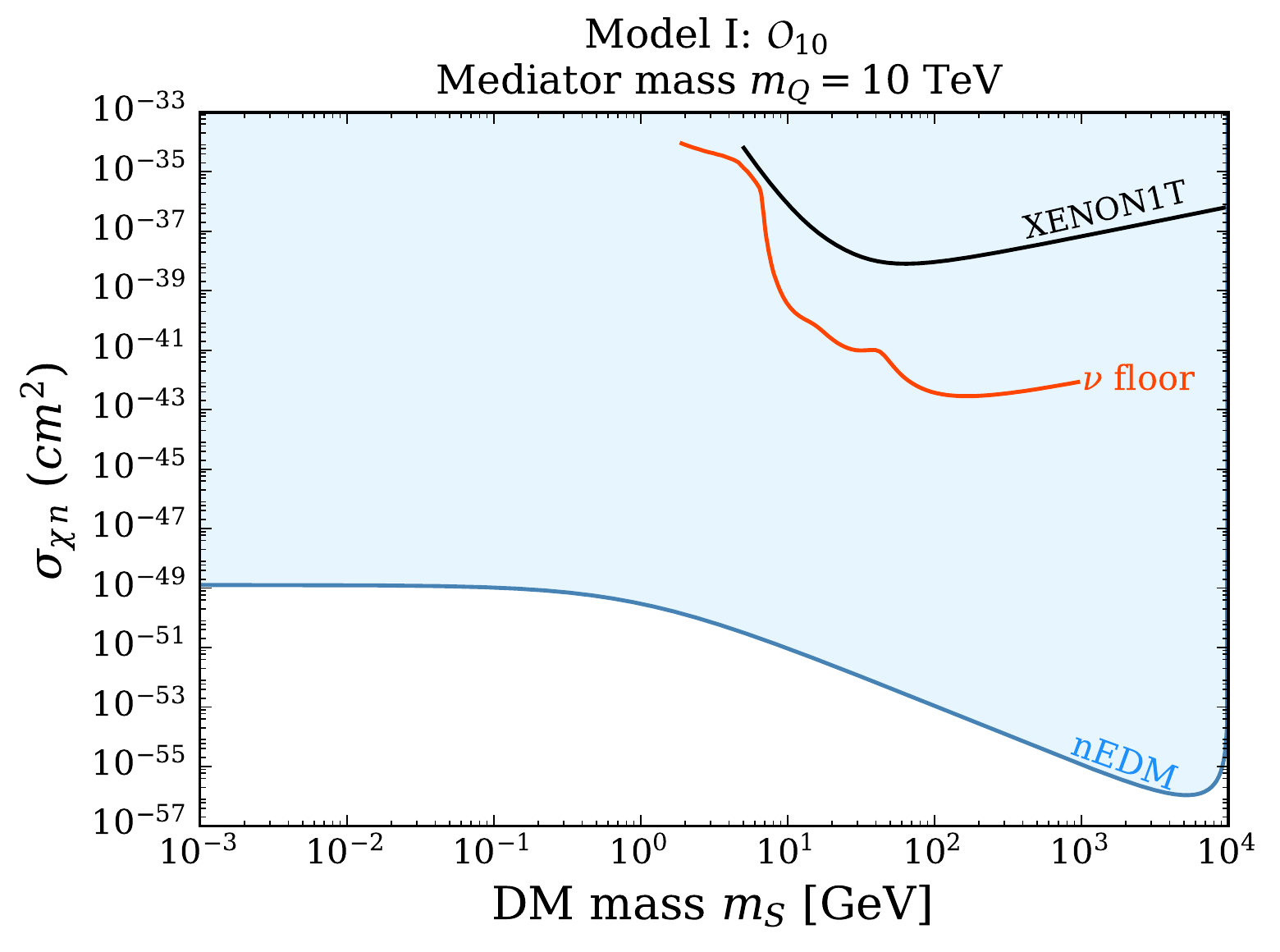}
\hfill
\includegraphics[width=0.45\linewidth]{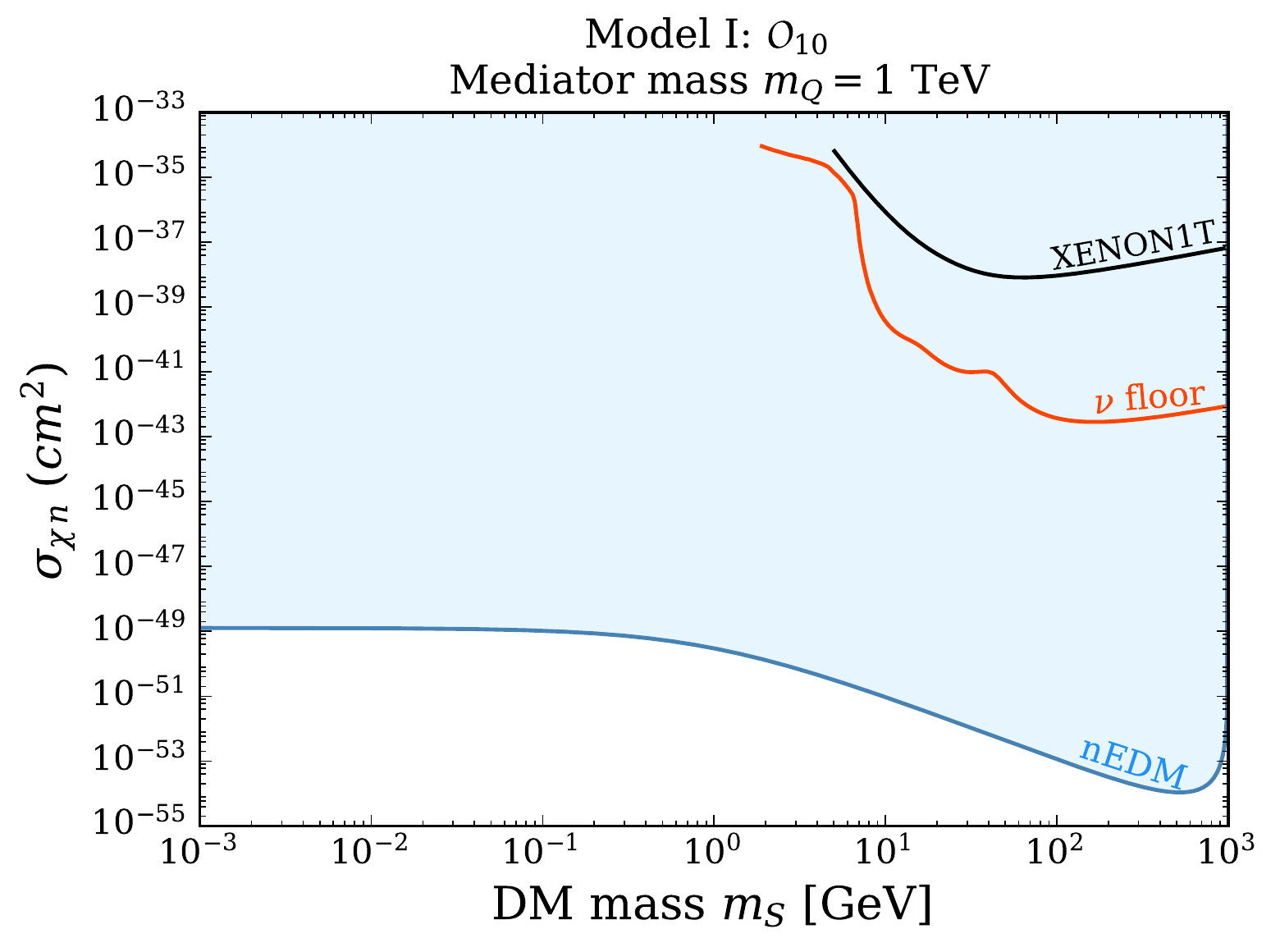}
\includegraphics[width=0.45\linewidth]{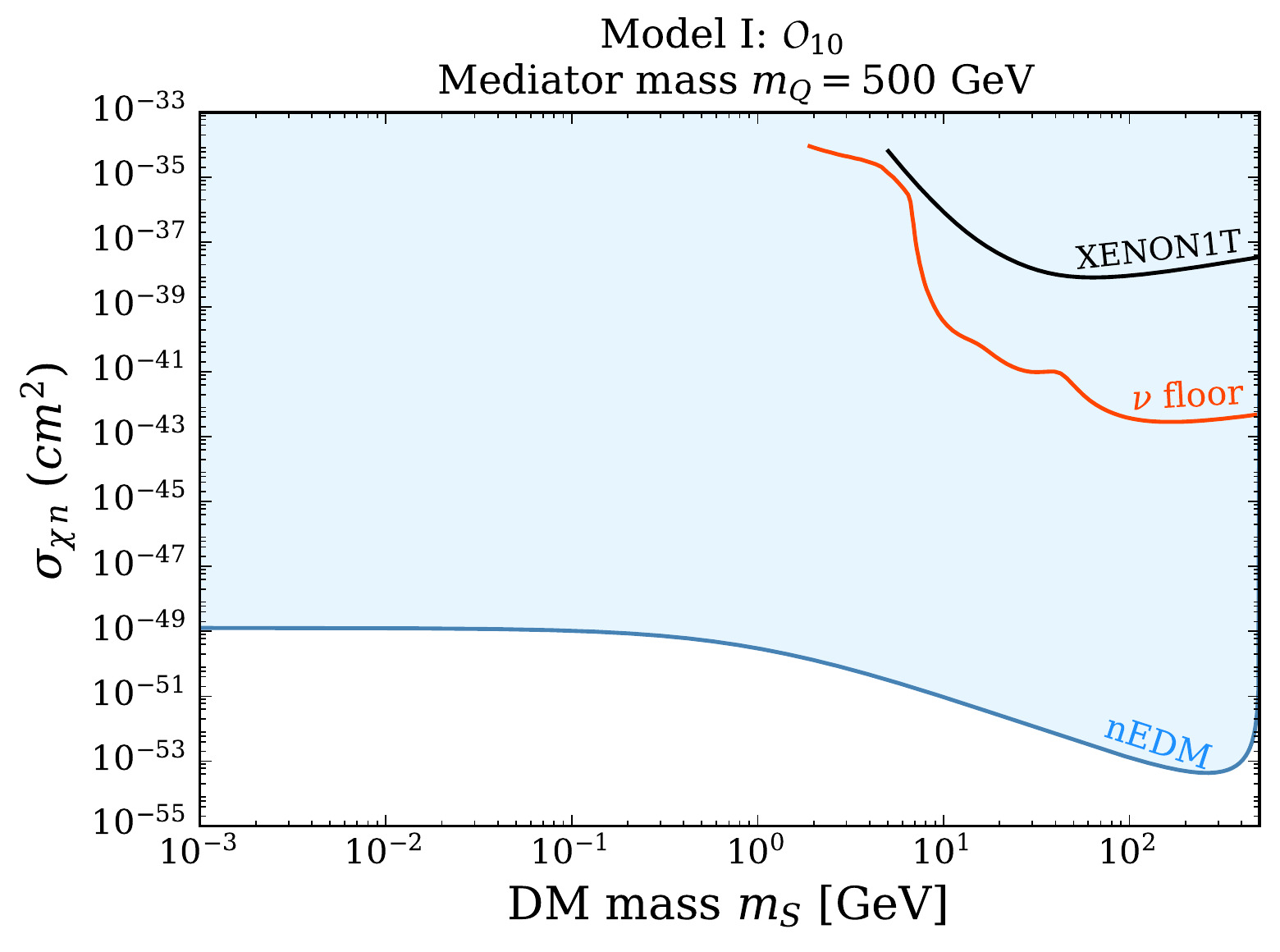}
\hfill
\includegraphics[width=0.45\linewidth]{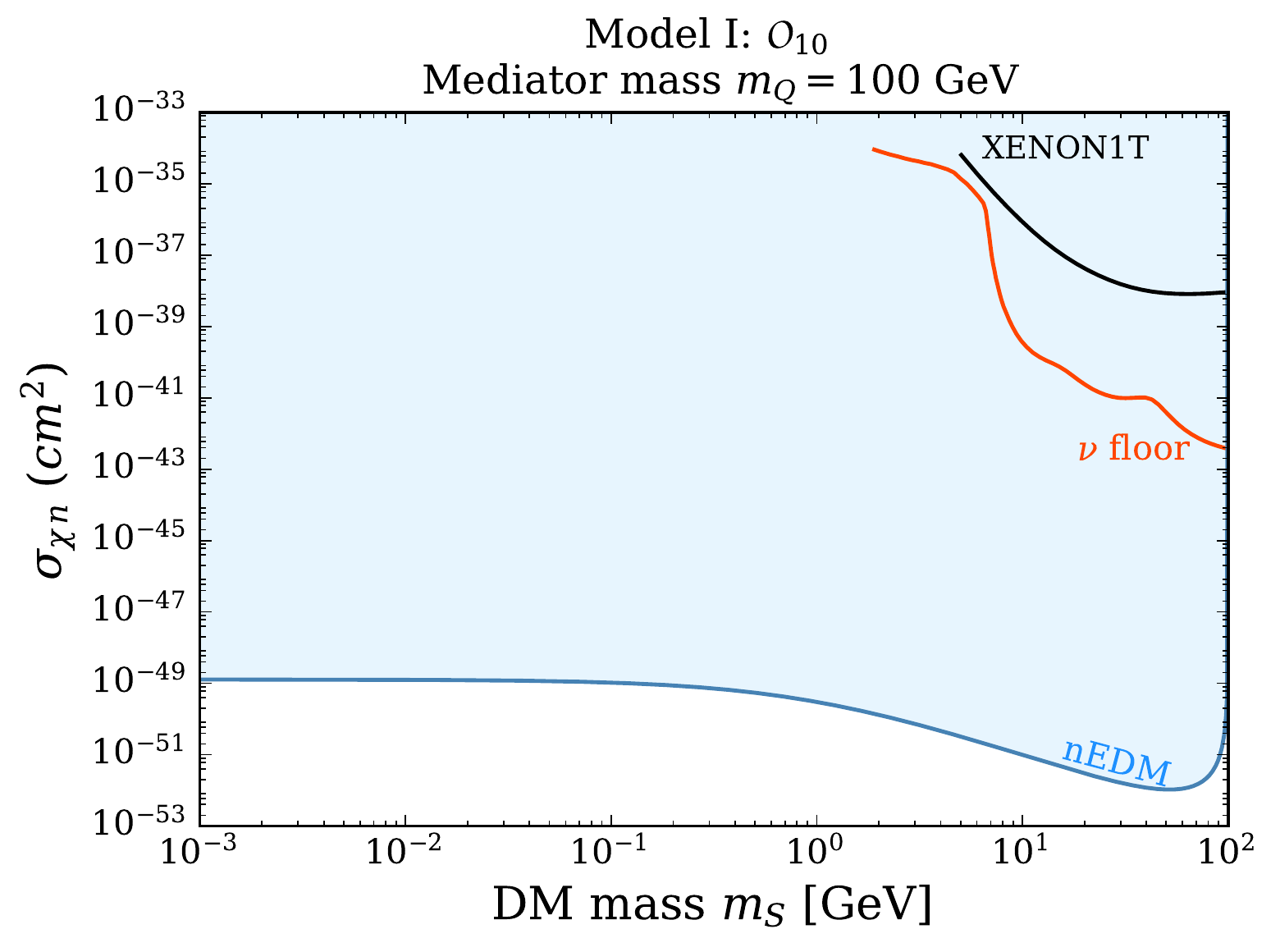}
\caption{The black curves show the current XENON1T upper bound on the
  DM-nucleon cross section for $ \Oten $ in Model I as a function of
  DM mass for four mediator masses, $m_Q = 0.1,0.5, 1$ and $10
  \TeV$. The blue shaded region is excluded by current limits on the
  nEDM. The orange curve denotes the neutrino floor for $ \Oten $ and
  has been taken from Ref.~\cite{Dent:2016iht}.}
\label{fig:Exc1}
\end{figure*}
	
We had seen in the previous Section that the same combination of
Yukawa couplings appears in the expression of the nEDM and in the
coefficients of the P- and T-odd operators contributing to
WIMP-nucleon interactions. The upper bound on the nEDM
\cite{Afach:2015sja} therefore leads to upper bounds on the
WIMP-nucleon scattering cross sections due to these operators. In this
section we discuss these constraints quantitatively, focusing on
$ \Oten $ for Model~I and $ \Oeleven $ for Model~II which are the
leading new operators (beyond $\Oone$ and $\Ofour$) in these
models. We compare the resulting constraints with the most recent
XENON1T results \cite{Aprile:2018dbl} as well as with the irreducible
``neutrino floor'' background in Xenon experiments.

We compute the nEDM induced from quark EDM and CEDM dimension-5
contributions which were given in the previous Section. We assume that
these are the only contributions to the (color) EDMs of the quarks. In
order to calculate the value of the nEDM from quark EDM $d_q$ and CEDM
$\tilde d_q$, we use the following formula:
\begin{equation} \label{eq:nEDM_lattice_sumrules}
  d_n = g^u_T d_u +  g^d_T d_d + g^s_T d_s + 1.1 \, e \,
  ( 0.5 \, \tilde{d}_u + \tilde{d}_d )\,.
\end{equation}
Here the tensor charges $ g^u_T = -0.233(28) $, $ g^d_T = 0.774(66) $
and $ g^s_T = 0.009(8) $ have been calculated using lattice QCD
\cite{Bhattacharya:2015esa,Bhattacharya:2015wna} (see also
Refs.\cite{Gupta:2018lvp,Yamanaka:2018uud,Alexandrou:2017qyt}) at a
renormalization scale of $2 \GeV$. We are not aware of a reliable
lattice computation of the contribution of the chromo-EDMs to $d_n$;
we therefore employ a computation using QCD sum rules evaluated at a
renormalization scale of $2 \GeV$ \cite{Pospelov:2000bw}, although
there is an $ \mathcal{O}(50 \%) $ uncertainty in these results
\cite{Hisano:2012sc,Fuyuto:2012yf}. We will again assume that the new
Yukawa couplings are universal. In this case the uncertainty in the
coefficients in Eq.(\ref{eq:nEDM_lattice_sumrules}) might shift the
boundary of the excluded region slightly, without affecting our
results qualitatively. We also stress that these results are
independent of the cancellation relation (\ref{eq:forbid_O1_Model_I})
or (\ref{eq:forbid_O1_Model_II}) that one has to impose if the
contribution of the new operators is to be significant.

Using the computed expressions for quark EDM and CEDM in
Eq.\eqref{eq:qEDM_Model_I} and \eqref{eq:qEDM_Model_II}, we find
limits on $ | \text{Im}(y_1 y_2^\dagger) | $ and
$ | \text{Im}(l_1 l_2^\dagger) | $ respectively according to the
stringent experimental bound \cite{Afach:2015sja} given in
Eq.\eqref{eq:nEDM_limit}. We then convert these limits on the
imaginary part of the product of Yukawa couplings to limits on the
cross section, using the relations given in Tables
\ref{table:coeff_model_i} and \ref{tab:Model_II_Op_Coeff} as well as
the following expression for the ``WIMP-nucleon scattering cross 
section"\footnote{These quantities can be obtained by setting momentum 
transfer and WIMP velocity to unity and are used to facilitate comparison of 
the Wilson coefficients for elastic scattering on different target nuclei 
across all NREFT operators. However, the physical cross sections for 
sub-leading operators contain appropriate factors of
momentum transfer and/or WIMP velocity.}\cite{DeSimone:2016fbz}:
\begin{equation} \label{eq:Cross_Section_Converter}
  \sigma_{\Oeleven} = \dfrac{\mu_{\chi N}^2}{\pi} \, (c_{11}^N)^2 \quad \text{and}
  \quad \sigma_{\Oten} = \dfrac{3 \mu_{\chi N}^2}{\pi} \, (c_{10}^N)^2 \; .
\end{equation}
We note that due to the chosen normalization of the operators, these expressions
look very similar to those for the traditional operators $\Oone$ and $\Ofour$.

In order to compare the nEDM-derived 90\% c.l. limit with the XENON1T
sensitivity, we assume a standard isothermal DM halo with $ \rho_\chi
= 0.3 \GeV \text{cm}^{-3}$, $ v_0 = 220 $ km/s, $ v_e = 232 $ km/s
and $ v_{esc} = 544 $ km/s. We use the Mathematica code
\texttt{dmformfactor} \cite{Anand:2013yka} for computing the Xenon
nuclear response functions $ W_k^{NN'} $ but use our own routines to
calculate the exclusion limits using the procedure outlined in
Appendix \ref{app:Xenon1T}. This leads to the current exclusion limits
on the WIMP-nucleon cross section for $ \Oten $ and $ \Oeleven $ at 90
\% C.L. from the most recent XENON1T results \cite{Aprile:2018dbl}. We
also show the irreducible background level from coherent
neutrino-nucleus scattering (``neutrino floor'') as estimated in
Ref.~\cite{Dent:2016iht}

Figure \ref{fig:Exc1} depicts the $ \sigma_{\rchi n} - m_S$ exclusion
involving the operator $ \Oten $ for Model I with four values of the
common mediator mass $m_Q$. We see that the current XENON1T exclusion
contour does not go below $10^{-2}$ pb.  Since $ \Oten $ is a
momentum-suppressed SD operator, the limits plotted are roughly
$ q^2/(m_N^2 A^2) \sim \mathcal{O}(10^{-8}-10^{-9}) $ suppressed in
comparison to the SI results quoted by the XENON1T collaboration.

In order to derive the constraint on the scattering cross section that
results from the bound on the nEDM, one can replace the combination
${\rm Im}(y_1 y_2^\dagger) m_Q$ by a constant (proportional to the
nEDM) divided by the loop function $F$ of
eq.(\ref{eq:Loop_Function_Model_I}). For $m_S^2 \ll m_Q^2$ the result
then scales like $m_N^2 / (m_S+m_N)^2$, i.e.  it approaches a constant
for $m_S \ll m_N \simeq 1$ GeV but scales like $1/m_S^2$ for $m_S > 1$
GeV. Note that this constraint is independent of $m_Q$; as long as
$m_S^2 \ll m_Q^2$ the four frames of Fig.~\ref{fig:Exc1} only differ in
the end point of the $x$ axis, which is given by $m_Q$. For
$m_S \simeq m_Q$ the constraint becomes somewhat weaker, partly
because the loop function $F$ becomes smaller, as we saw above, but
mostly due to the $1/(m_Q^2 - m_S^2)^2$ factor from the squared $Q$
propagator in the scattering cross section. We see that this indirect
constraint is more than twelve orders of magnitude below the present
XENON1T sensitivity, and at least eight orders of magnitude below the
neutrino floor. In other words, in this scenario the nEDM constraint
implies that the contribution from $\Oten$ to WIMP-nucleon
scattering is totally negligible.

\begin{figure*}
\centering
\includegraphics[width=0.45\linewidth]{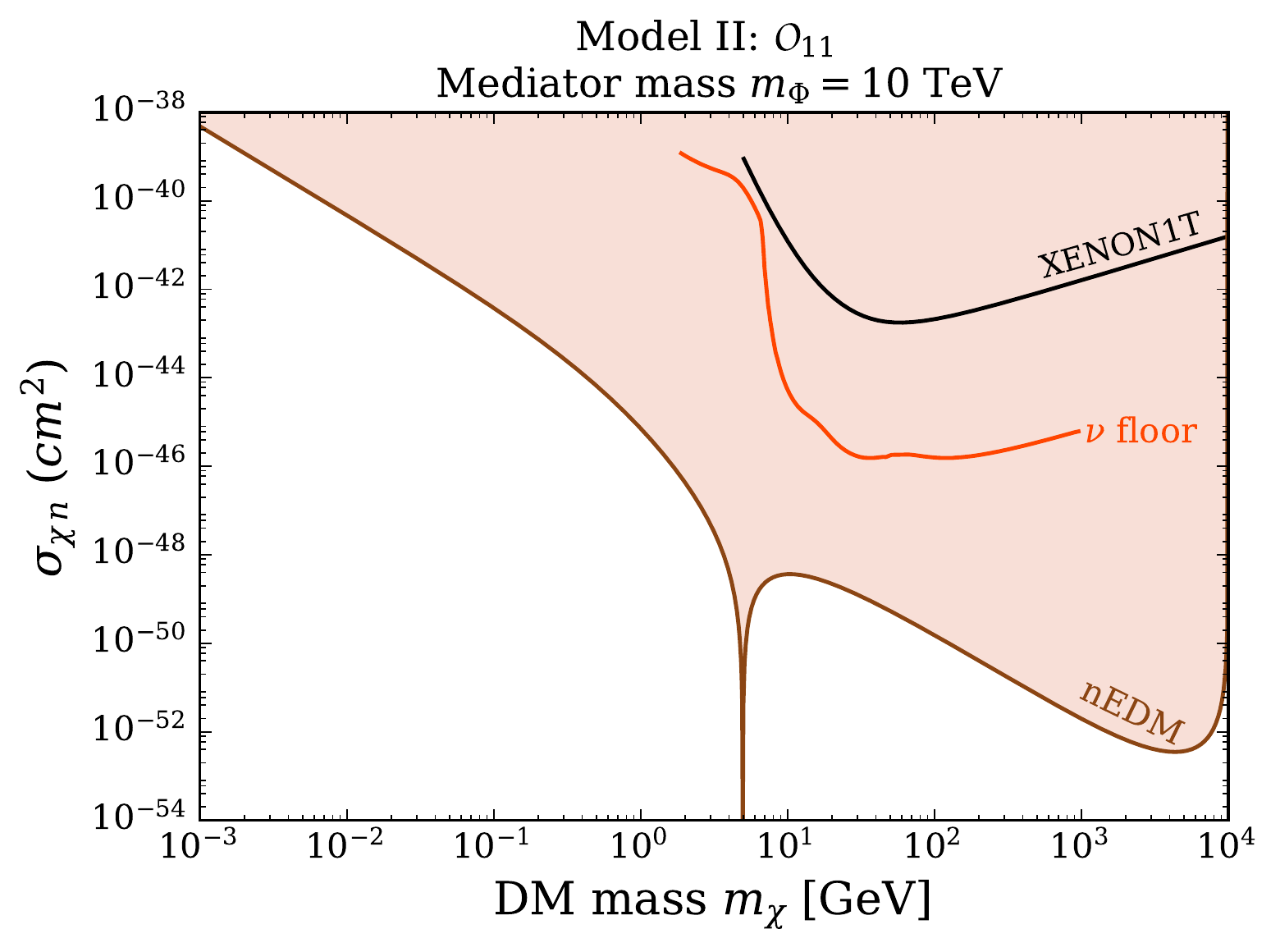}
\hfill
\includegraphics[width=0.45\linewidth]{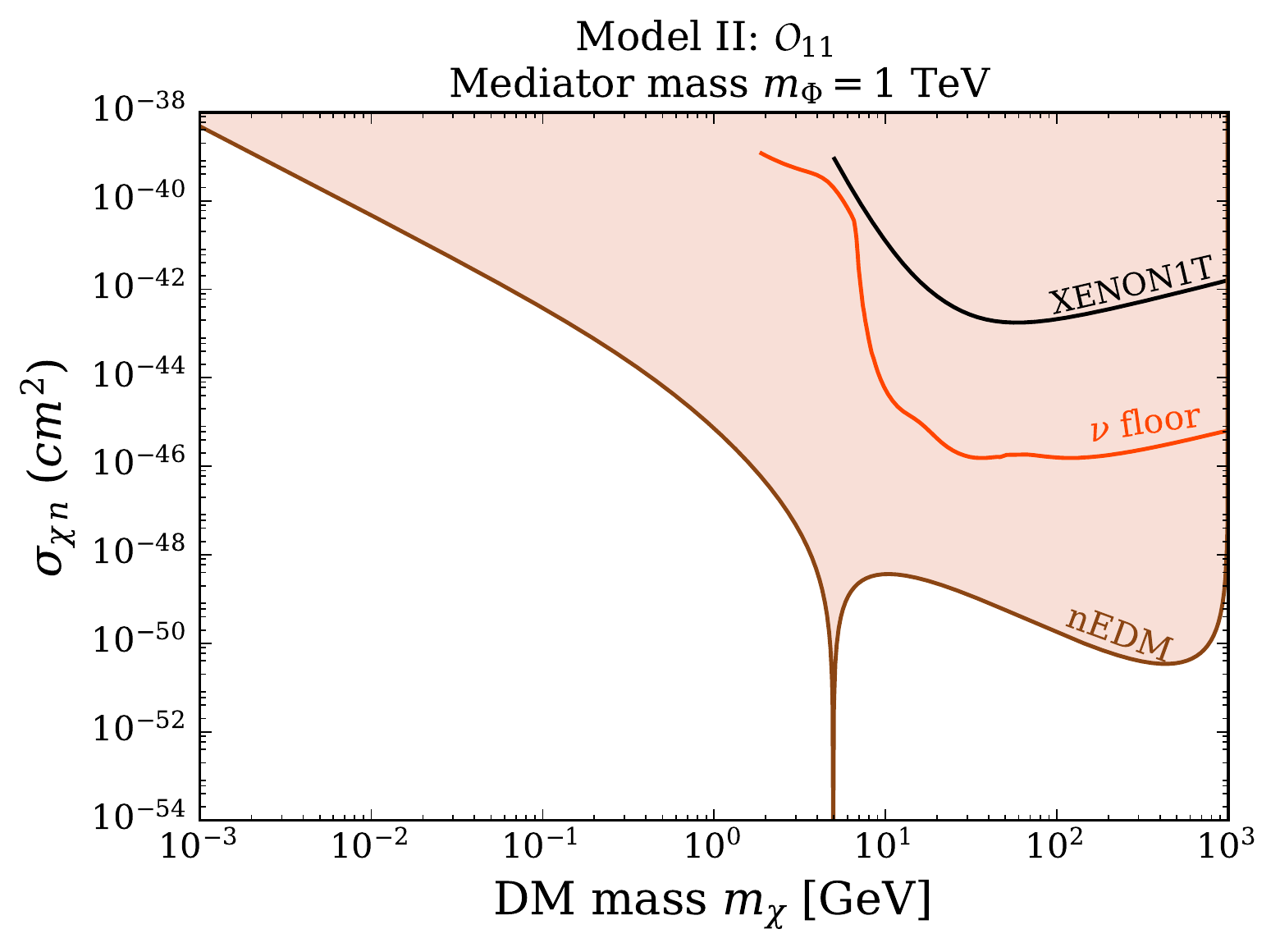}
\includegraphics[width=0.45\linewidth]{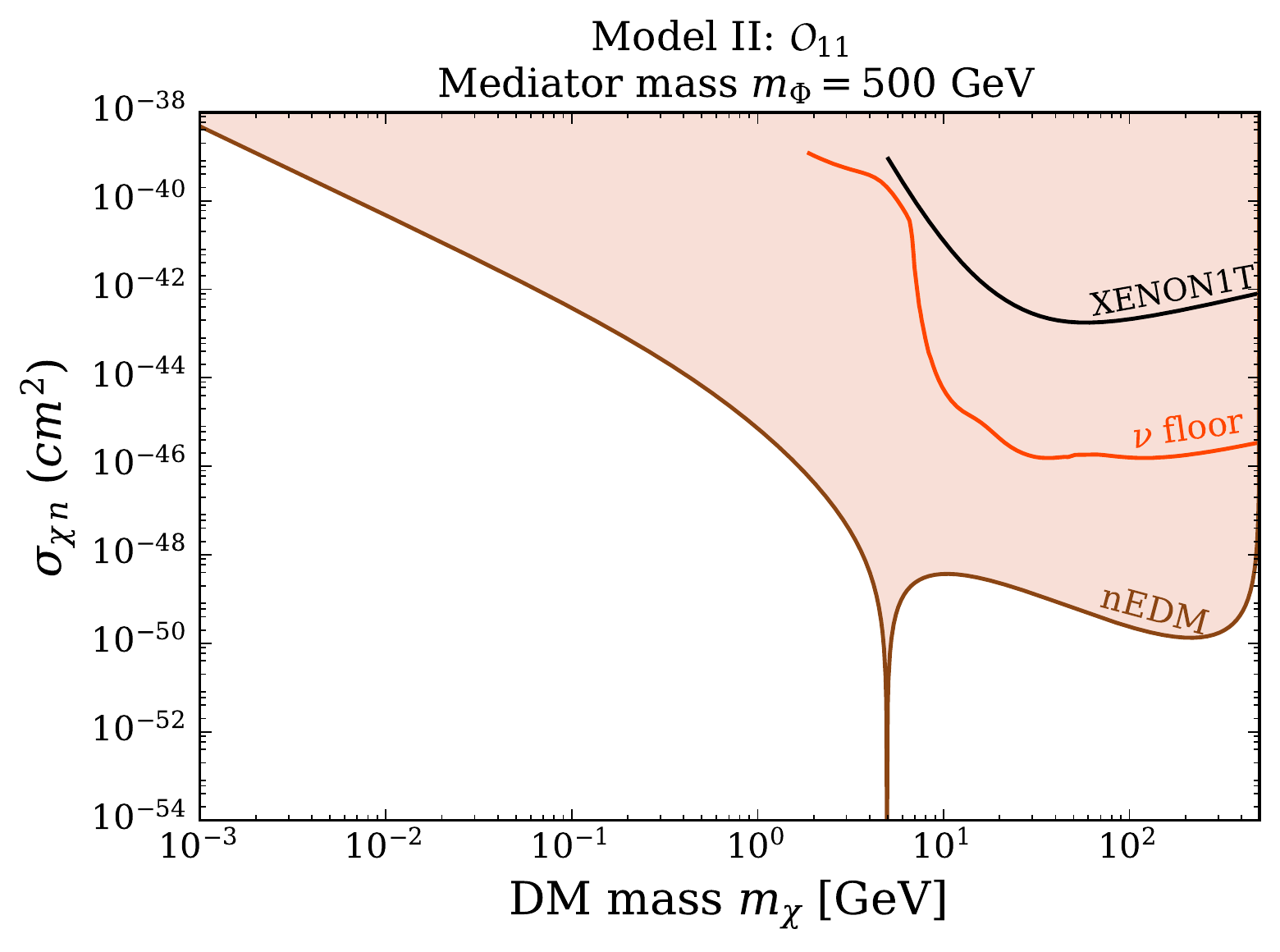}
\hfill
\includegraphics[width=0.45\linewidth]{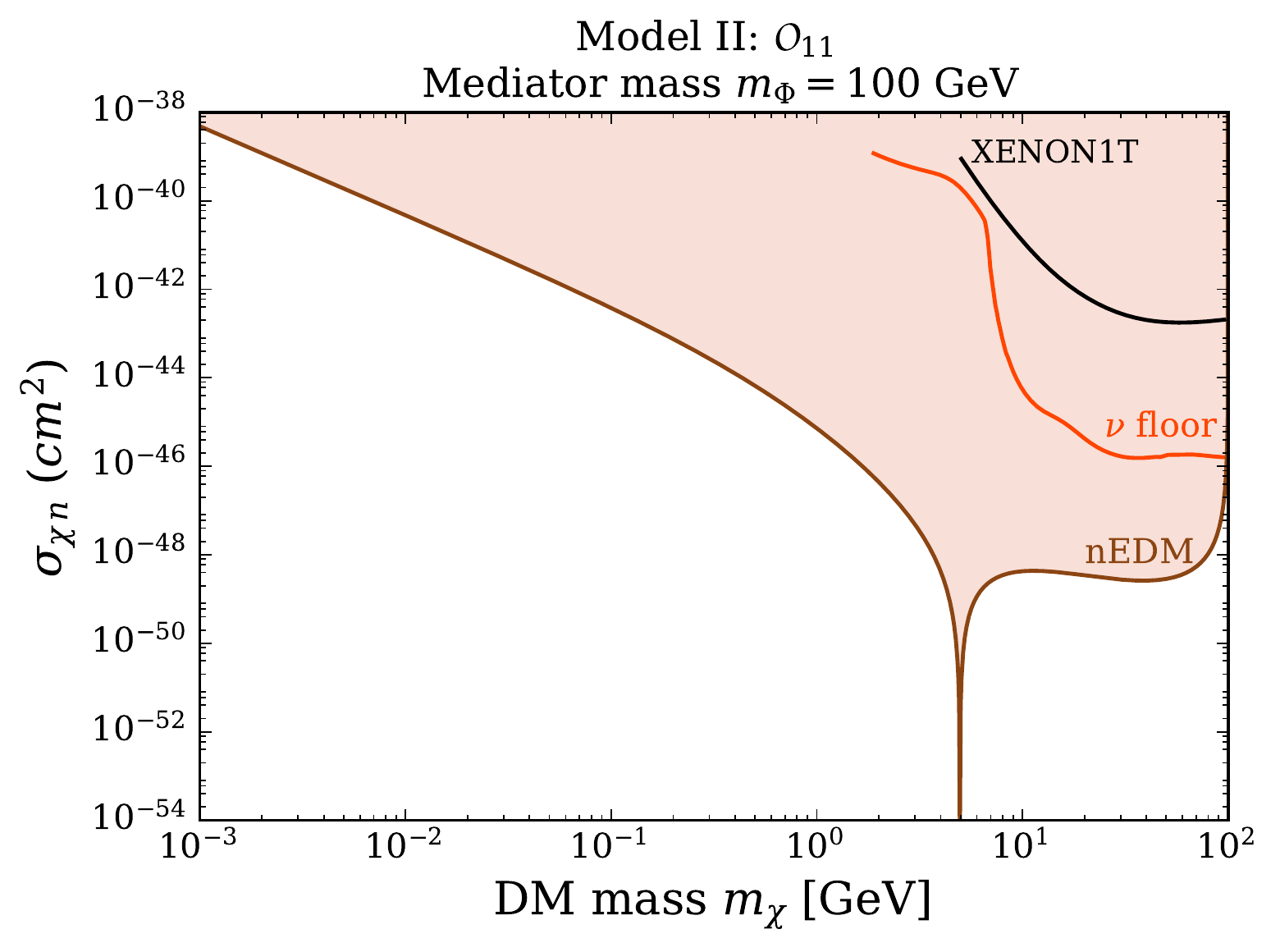}
\caption{The black curves show the current XENON1T upper bound on the
  DM-nucleon cross section for $ \Oeleven $ in Model II as a function
  of DM mass for four mediator masses $m_\Phi = 0.1,0.5, 1$ and
  $10 \TeV$. The shaded region is excluded at 90\% cl. by current
  limits on the nEDM. The orange curve denotes the neutrino floor for
  $ \Oeleven $ and has been taken from Ref.~\cite{Dent:2016iht}.}
\label{fig:Exc2}
\end{figure*}

Figure \ref{fig:Exc2} shows analogous results for the $ \Oeleven $
operator in Model II. $ \Oeleven $ is independent of the nuclear spin
but momentum suppressed; in fact, for small $q = |\vec{q}|$ the
suppression must be of order $q^2/m_A^2$ rather than $q^2/m_N^2$ since
a pointlike (unresolved) nucleus does not ``know'' about its nucleonic
constituents. The XENON1T limit on the cross section is therefore
weaker by three to four orders of magnitude compared to the case of
the traditional SI operator $\Oone$.

Regarding the indirect constraint on the scattering cross section from
the nEDM bound, the situation is slightly more complicated here than
in Model~I because according to Table~\ref{tab:Model_II_Op_Coeff}
there are two contributions to the Wilson coefficient $c_{11}$, with
different dependence on the WIMP mass. The constraint can be derived
by replacing ${\rm Im}(l_1l_2^\dagger)$ by a constant (again involving
$d_n$) divided by the WIMP mass and the loop function $G$. As long as
$m_\chi^2 \ll m_\Phi^2$ the nEDM constraint on
${\rm Im}(l_1l_2^\dagger) / m_\Phi^2$ thus scales like $1/m_\chi$. Let
us consider the cases $m_\chi \ll m_N$ and $m_\chi \gg m_N$
separately. In the first case $c_{11}$ is dominated by the
contribution $\propto m_N/m_\chi$ given in the third line of
Table~\ref{tab:Model_II_Op_Coeff}. The constraint on $|c_{11}|^2$ then
scales like $1/(m^4_\chi)$ and $\mu_{\chi N}^2 \simeq m_\chi^2$, hence
the constraint on the scattering cross section varies
$\propto 1/m_\chi^2$. For $m_\chi \gg m_N$, $c_{11}$ is dominated by
the contribution from the last line in
Table~\ref{tab:Model_II_Op_Coeff}, i.e. the nEDM constraint on
$|c_{11}|^2$ scales like $1/m_\chi^2$. Since now
$\mu_{\chi N} \simeq m_N$ the nEDM constraint on the cross section
again scales $\propto 1/m_\chi^2$.  Since the two contributions to
$c_{11}$ have opposite sign, there is a cancellation for
$m_\chi \sim m_N$, leading to a very strong upper bound on the cross
section.

For $m_\chi \simeq m_\Phi$ the nEDM constraint becomes a bit weaker
again, chiefly due to the $1/(m_\Phi^2 - m_\chi^2)^2$ factor from the
squared mediator propagator. Altogether we nevertheless see that the
nEDM constraint is again several orders of magnitude below the
neutrino floor, i.e. the contribution from $\Oeleven$ to the
WIMP-nucleon cross section can be neglected.

\begin{figure*}
\centering
\includegraphics[width=0.45\linewidth]{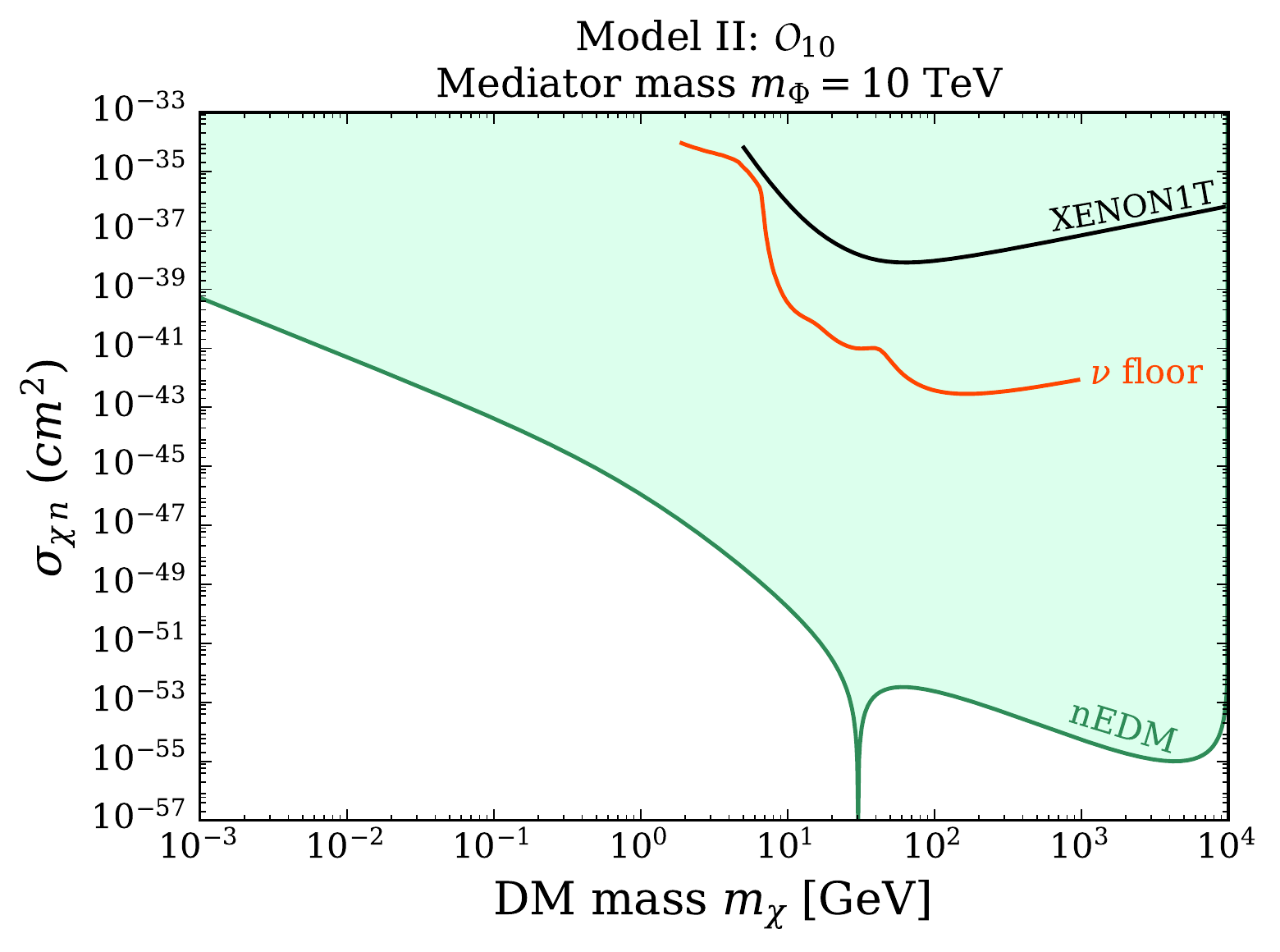}
\hfill
\includegraphics[width=0.45\linewidth]{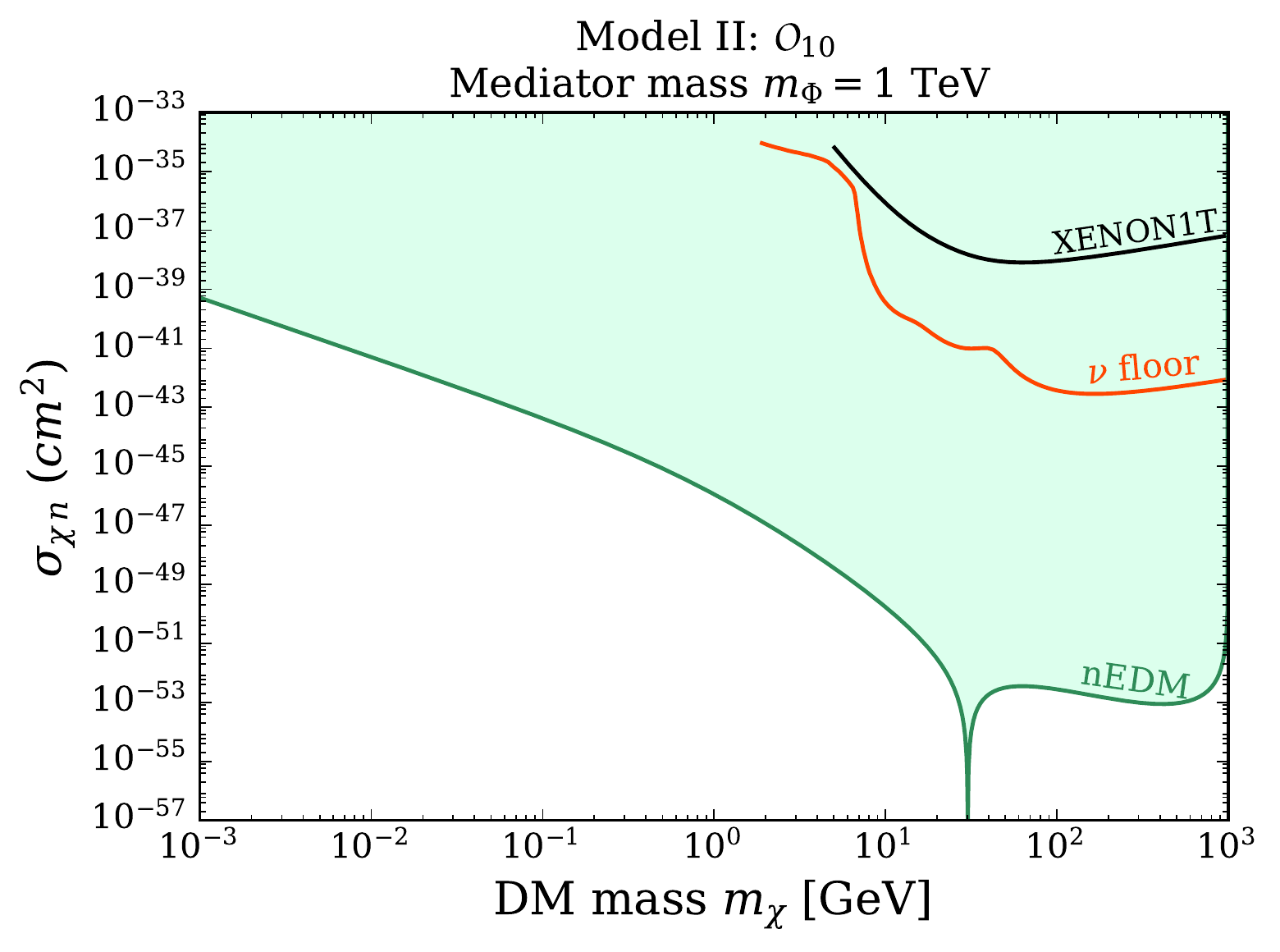}
\includegraphics[width=0.45\linewidth]{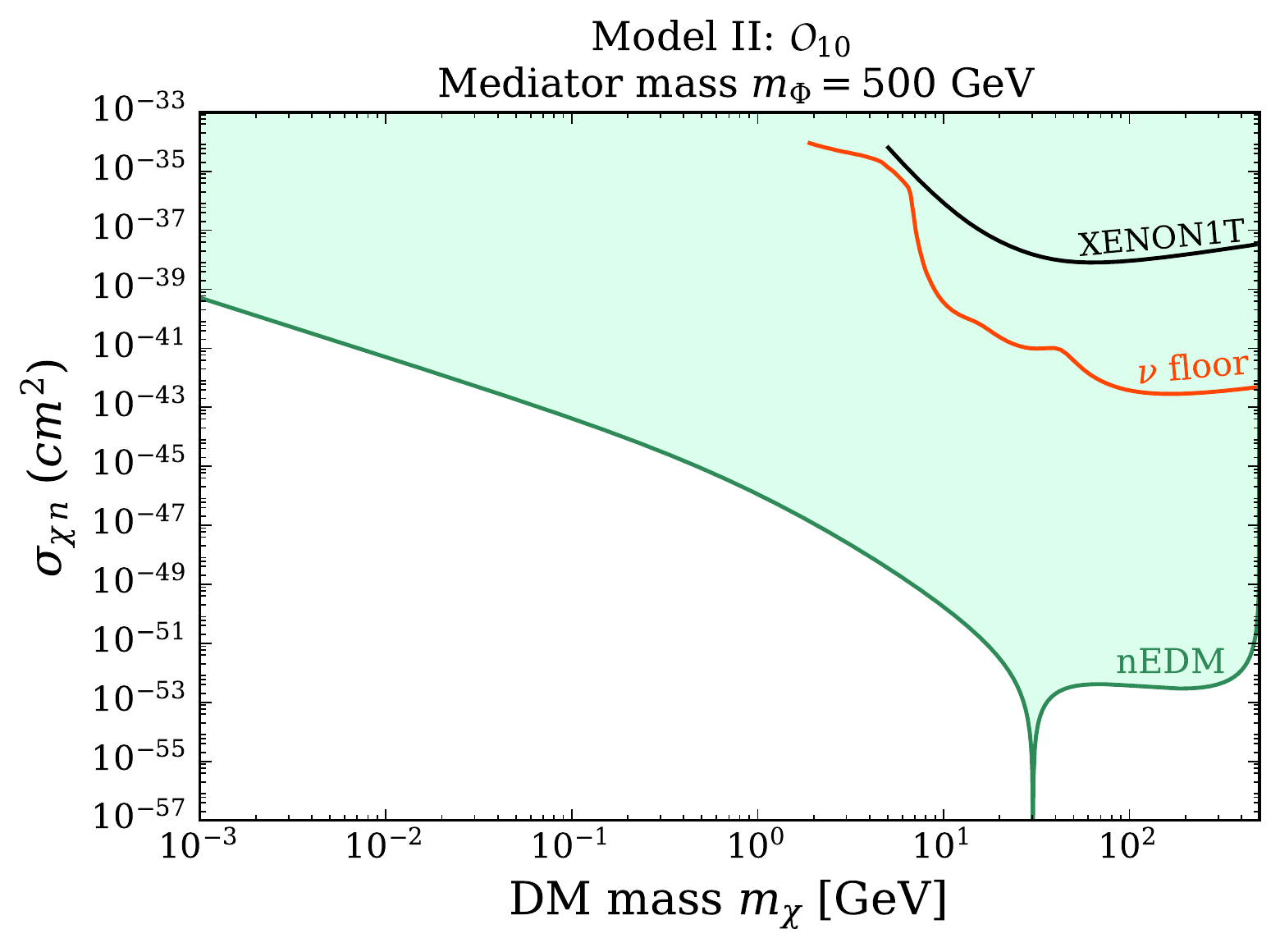}
\hfill
\includegraphics[width=0.45\linewidth]{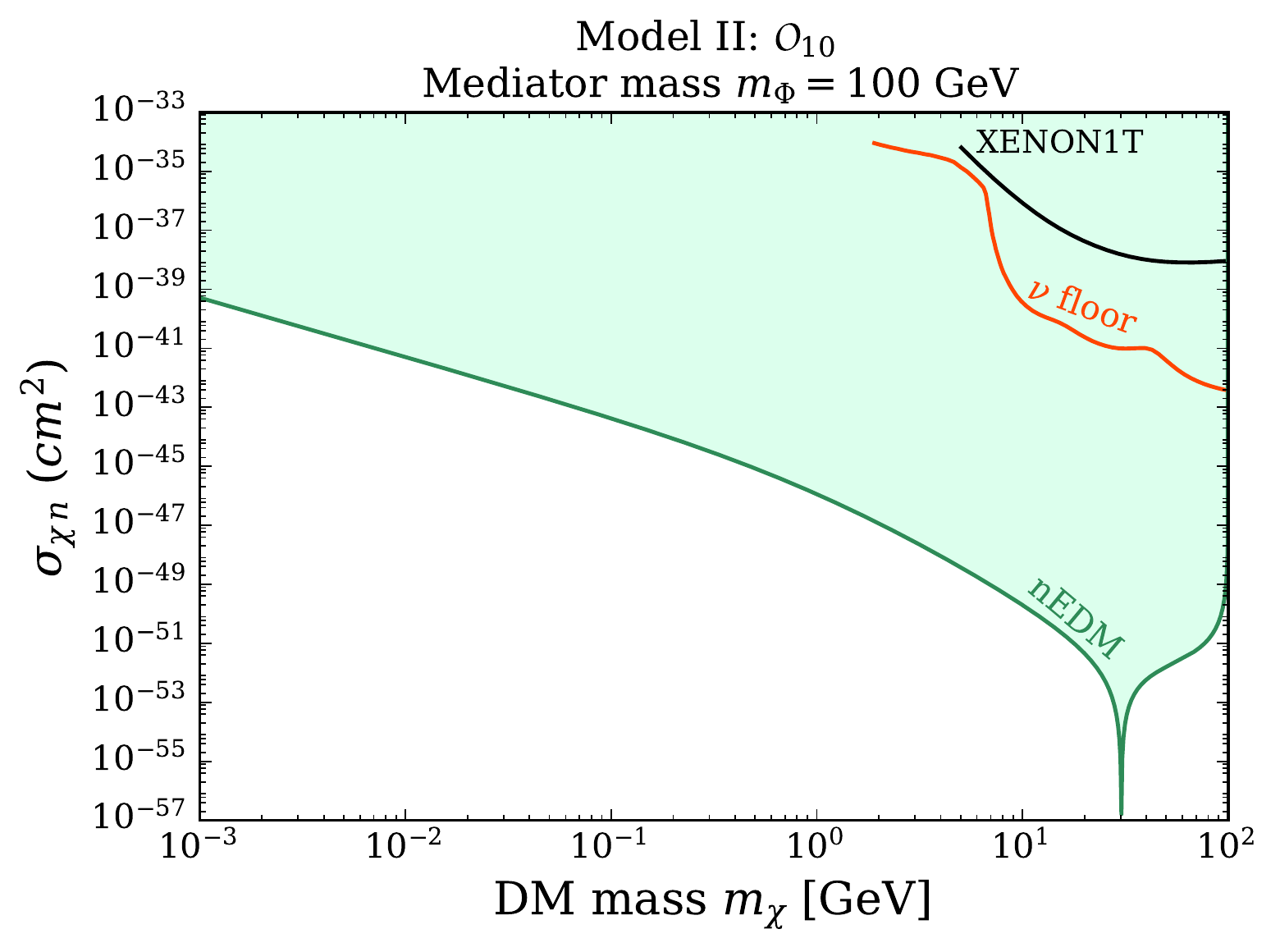}
\includegraphics[width=0.45\linewidth]{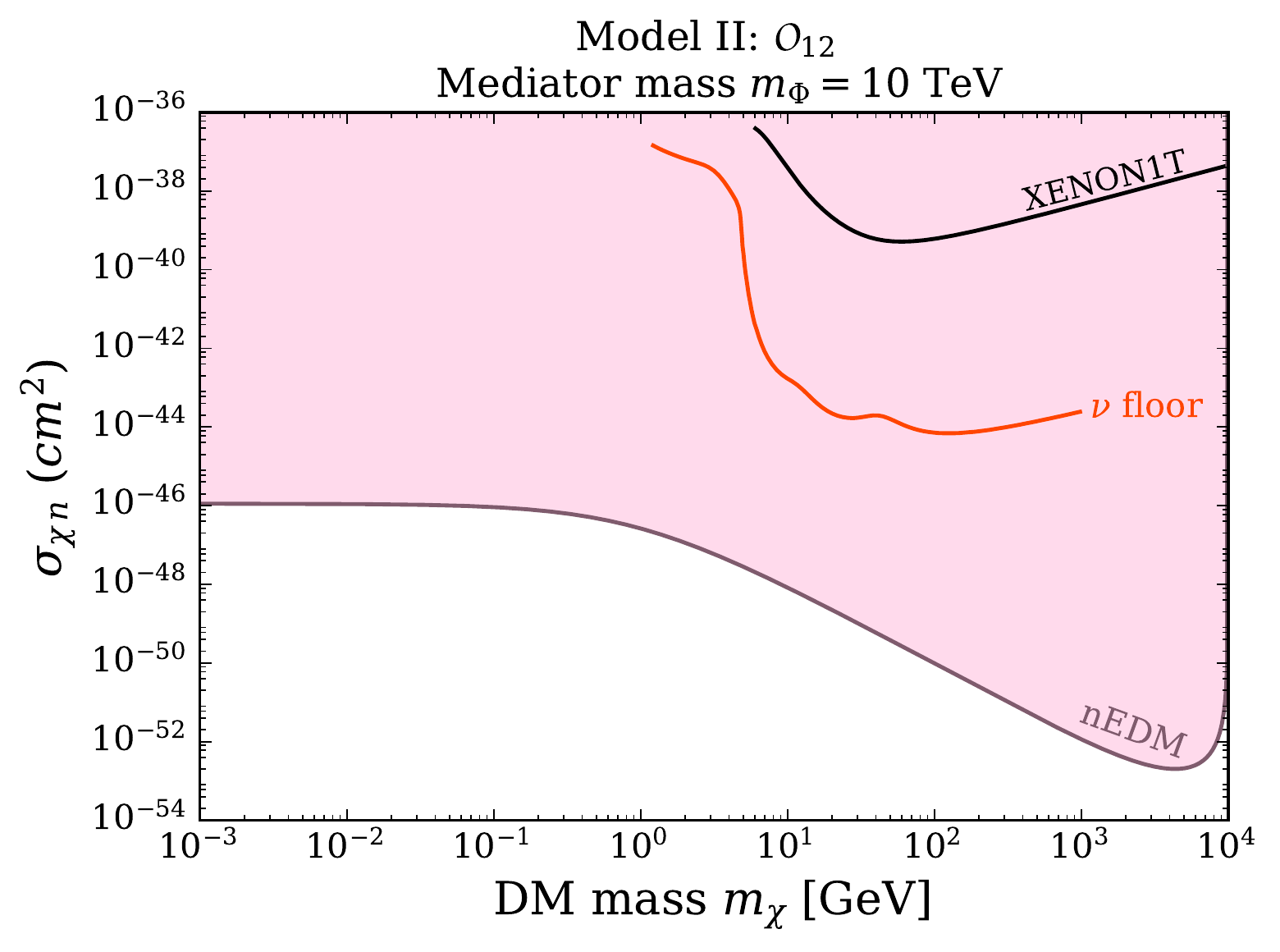}
\hfill
\includegraphics[width=0.45\linewidth]{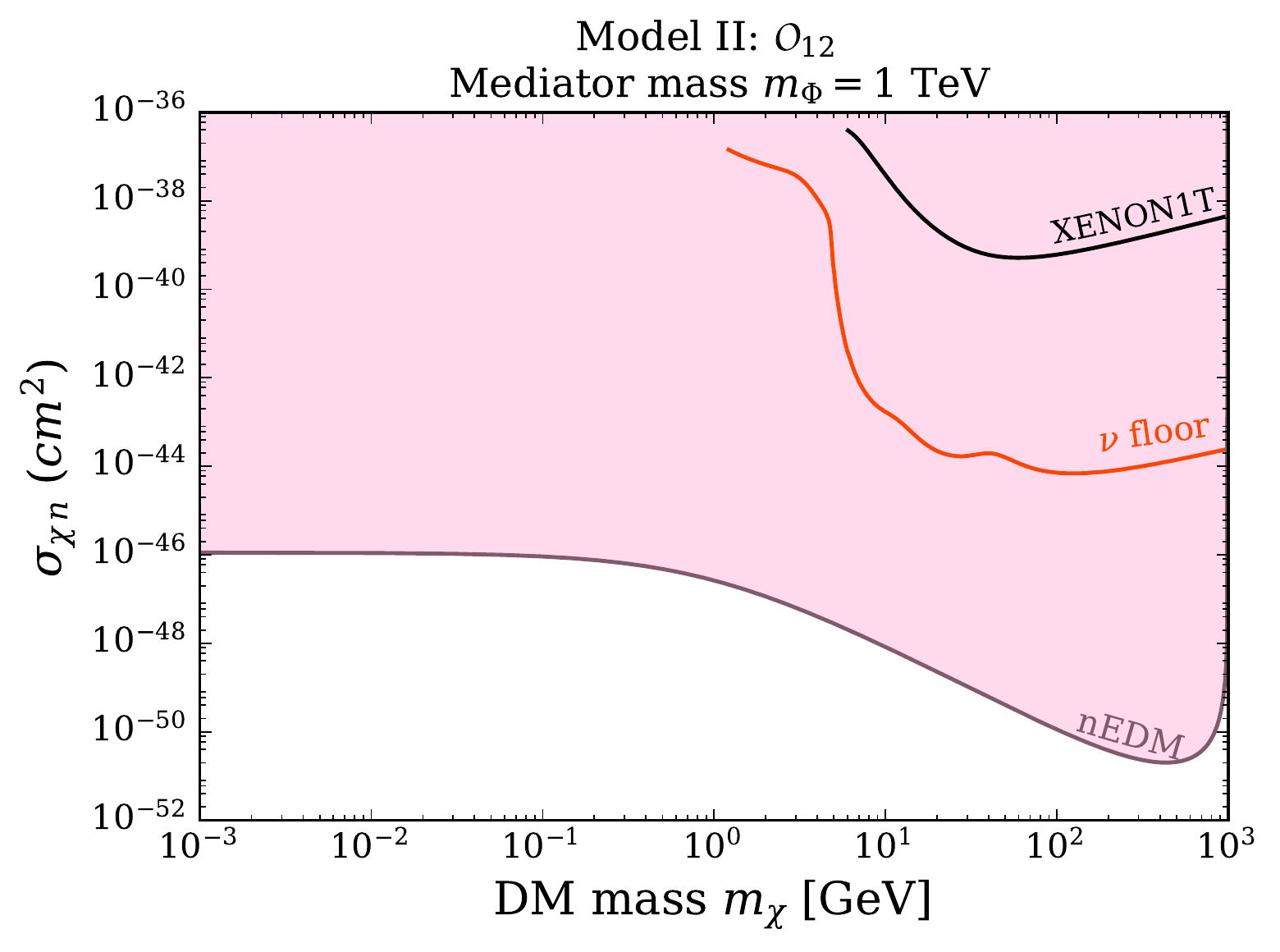}
\includegraphics[width=0.45\linewidth]{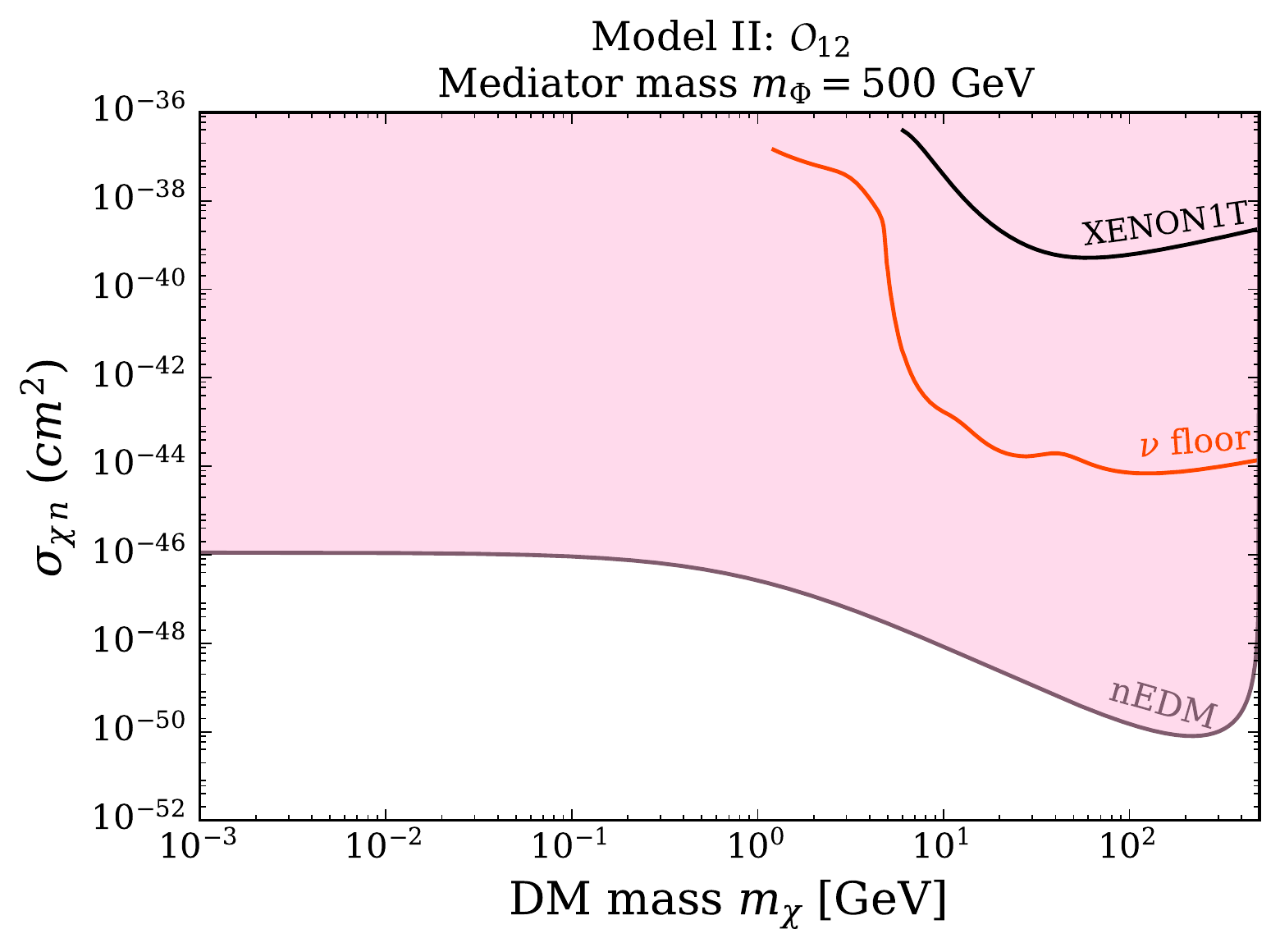}
\hfill
\includegraphics[width=0.45\linewidth]{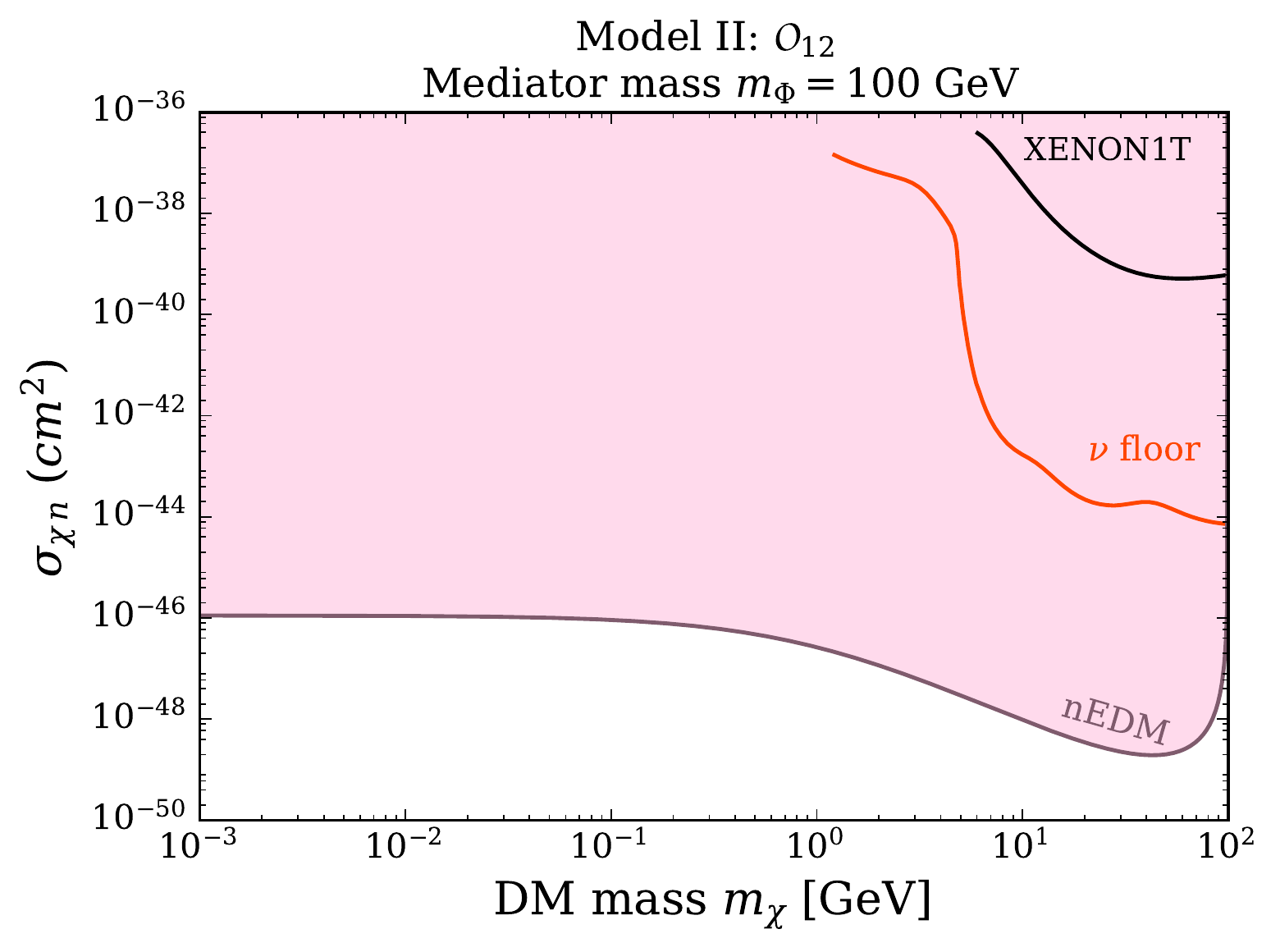}
\caption{DM-nucleon cross section for $ \Oten $ (the top four
  subfigures) and $ \mathcal{O}_{12} $ (the bottom four subfigures) in
  Model II as a function of DM mass for four values of the mediator
  mass, $ m_\Phi $ = 0.1, 0.5, 1 and 10 TeV. The shaded regions in
  green and pink denote the regions excluded by current limits on the
  nEDM. The orange curves denoting the neutrino floor for $ \Oten $
  and $ \mathcal{O}_{12} $ have been taken from
  Ref. \cite{Dent:2016iht}.}
\label{fig:Exc3}
\end{figure*}
              
Although we have shown above that for Model II, $ \Oten $ and
$ \mathcal{O}_{12} $ are expected to be sub-leading due to the
presence of $ \Ofour $ and $ \Oeleven $, in Fig.~\ref{fig:Exc3} we
repeat the exercise of Fig.~\ref{fig:Exc2}, but now for $\Oten$ (top
frames) and $\mathcal{O}_{12}$ (bottom frames). The current XENON1T
limit as well as the neutrino floor for $\Oten$ are quite similar to
those for $\Oten$ in Model~I, see Fig.~\ref{fig:Exc1}.
Table~\ref{tab:Model_II_Op_Coeff} shows that in Model~II the
coefficient of $\Oten$ also receives two contributions, one of which
scales like $m_N/m_\chi$; hence the dependence of the nEDM constraint
on $m_\chi$ is similar to that in Fig.~\ref{fig:Exc2}. However, since
the contribution that is not proportional to $m_N/m_\chi$ has a much
smaller coefficient, the cancellation now happens at significantly
larger WIMP mass than in Fig.~\ref{fig:Exc2}.  On the other hand,
$\mathcal{O}_{12}$ only receives one contribution, which is
independent of $m_\chi$ as long as $m_\chi^2 \ll m_\Phi^2$. The shape
of the nEDM constraint on the scattering cross section therefore
resembles that for Model~I shown in Fig.~\ref{fig:Exc1}. Evidently in
all cases the nEDM constraint once again implies that the
contributions from these P- and T-odd operators to WIMP-nucleon
scattering can safely be neglected.

\section{Conclusions} \label{sec:outlook}
	
We studied the detection prospects of P- and T-odd operators arising
in the NREFT formalism for WIMP-nucleus elastic scattering. These
operators occur only if one includes terms that are suppressed by
powers of relative velocity or three-momentum transfer. They can therefore
be expected to be significant only if the coefficient of the
leading spin-independent operator $\Oone$ is very small or vanishes.
Moreover, T-odd NREFT operators can only arise in the low-energy
limit of a relativistic theory if the latter violates CP invariance.
This can give rise to very stringent constraints on the theory,
in particular from electric dipole moments.

We analyzed these issues in the framework of two simplified models,
taken from ref.~\cite{Dent:2015zpa}. They introduce one or more
$s-$channel mediator(s) carrying electric and color charge, with spin
$1/2$ (Model~I) or $0$ (Model~II); the dark matter particle then has
spin $0$ or $1/2$, respectively. These models will cause FCNC at
one-loop unless each mediator only couples to one quark. Moreover, we
pointed out that generically the coefficient of $\Oone$ is {\em not}
suppressed in these models. Since by now strong constraints on
spin-independent WIMP-nucleon scattering exist for several isotopes,
the new operators will be significant only if the coefficient of
$\Oone$ is suppressed for both neutrons and protons. This generically
requires {\em two independent} cancellations, which should hold to one
part in $10^3$ or better. We also note that one has to introduce {\em
  several mediators} for these cancellations to be possible at all,
i.e. in minimal versions of these models, with a single mediator, the
contribution from $\Oone$ will always dominate.

We therefore assumed that the new couplings and mediator masses are
universal for all generations. Since in this case the new terms in the
Lagrangian respect strong isospin invariance the two cancellation
conditions become almost the same. However, we showed that these
scenarios give rise to one-loop contributions to the electric dipole
moment of the neutron (nEDM) which are proportional to exactly the
{\em same} combination of new Yukawa couplings which appears in the
coefficients of the new P- and T-odd NREFT operators. The resulting
indirect constraint lies well below the irreducible neutrino
background. This implies that the contributions from the new T-odd
operators on the WIMP-nucleon scattering can safely be neglected in
the analyses of both current and future experiments. This conclusion
can only be avoided if one relaxes the universality assumption on the
new couplings and mediator masses. However, one will then have to
impose {\em three independent conditions}, in order to cancel the
coefficients of $\Oone$ for protons and neutrons as well as the
neutron EDM. Since these cancellations are not enforced by any
symmetry, we conclude that it is very unlikely indeed that the new
T-odd operators can contribute significantly. We emphasize that, for a
given size of the Wilson coefficients, these are the most important of
the total eleven new operators. This indicates that at least in models
with charged mediator coupling to light quarks, WIMP-nucleon
scattering can safely be analyzed using the traditional operators
$\Oone$ and $\Ofour$ only.

Charged mediators coupling exclusively to heavy ($c,b,t$) quarks will
contribute to WIMP-nucleon scattering only at one-loop level. On the
other hand, the neutron EDM constrains the EDMs of heavy quarks only
weakly; the nEDM may then receive dominant contributions at the
two-loop level, e.g. via Weinberg's three-gluon operator
\cite{Weinberg:1989dx}. In this case the contribution to the nEDM
would still only occur at one order higher in perturbation theory than
WIMP-nucleon scattering, as is the case in the scenarios we analyzed
here, so the conclusions are likely to be similar. Finally, the P- and
T-odd operators can also be generated from simplified models not
containing charged mediators \cite{Dent:2015zpa}. We leave the
analysis of these scenarios to future work.

\section*{Acknowledgments}

We thank Andreas Wirzba and Bastian Kubis for discussions.  RM was
partially supported by the Bonn Cologne Graduate School of Physics and
Astronomy.
	
\appendix
\section{Xenon1T limits} \label{app:Xenon1T}
\numberwithin{equation}{section}

The latest results from the XENON1T collaboration involved a run time
of 278.8 days with a target mass of 1.3 tonnes. To compute the limits,
we use here the data reported for `0.65t' of fiducial region in Table
I of \cite{Aprile:2018dbl}, which has a substantially smaller
background. Using the computed differential event rate in
\eqref{eq:diffrate}, we calculate the total number of expected signal
events $ N_s $ using the procedure outlined in \cite{Aprile:2011hx,
  Aprile:2015uzo}:
\begin{align}	\label{eq:X1T_events}
  N_s = \text{Exposure} \times \int_{S1_{\text{low}}}^{S1_{\text{upp}}}  \sum_{n=1}^{\infty} \,
  \text{Gauss}(S1|n, \sigma\sqrt{n}) \times \int_{0}^{\infty} \dfrac{dR}{dE_R}  \,
  \epsilon(E_R) \, \text{Poiss}(n|n_{ph}(E_R)) \; dS1 \, .
\end{align}
The differential recoil energy spectrum is converted to a differential
rate in the number of photoelectrons $ n $ by convoluting it with the
product of a Poisson distribution with mean $ n_{ph}(E_R) $, which is
the expected number of photoelectrons given a recoil energy $E_R$,
and the detection efficiency $\epsilon(E_R)$. We obtain
$n_{ph}(E_R)$ from the left panel of Fig.~3 of \cite{Aprile:2018dbl}
where the constant recoil energy contours intersect the middle of the
median nuclear recoil band and the $ 2 \sigma $ quantile curve. We
digitize the detection efficiency as a function of recoil energy from
the black curve of Fig.~1 of \cite{Aprile:2018dbl} and weigh it with a
factor of 0.5, since we only consider the fiducial region, and a factor of
0.475, since we only consider events in the reference region in
$ (\text{cS2}_b, \text{cS1}) $ as in Fig.~3 of \cite{Aprile:2018dbl}.

In order to finally obtain the event rate differential in the S1
signal, we apply to each photoelectron a convolution with a Gaussian
distribution of width $ \sigma = 0.4 \, \text{PE} $ which
parameterizes the response of the XENON1T photomultiplier tubes
\cite{Aprile:2015uzo}. We then integrate over the S1 signal region as
reported in \cite{Aprile:2018dbl} from $ S1_{\rm low} = 3 \, \text{PE} $
to $ S1_{\text{upp}} = 70 \, \text{PE} $ and multiply by the exposure
to get the total number of expected events $ N_s $. The experiment
observed $ N_{\text{s}} = 2 $ events with a best fit of number of
expected background events $ N_{\text{b}} = 0.83$ and therefore, we
employ a fairly simple Poisson upper limit on the number of signal
events at 90 \% C.L. after using \eqref{eq:X1T_events} to place
exclusions on different NREFT operators. Using this procedure, we
obtain satisfactory agreement (off by a factor of roughly 2) with the
XENON1T exclusion limit \cite{Aprile:2018dbl} for the SI response, which is
based on more sophisticated likelihood fits.
	
\section{Fierz Identities} \label{app:Fierz_Identities}

We use the following Fierz rearrangement identities as derived in
\cite{DeVries:1970xvk,Good:1955em} to derive the relativistic
effective operators for Model II.
\begin{align}
  (\bar{q}\chi)(\bar{\chi}q) &= -\frac{1}{4} \left[ \bar{q}q\bar{\chi}\chi
          + \bar{q}\gamma^{\mu}q\bar{\chi}\gamma_{\mu}\chi
          +\bar{q}\gamma^{5}q\bar{\chi}\gamma^{5}\chi
          -\bar{q}\gamma^{\mu}\gamma^{5}q\bar{\chi}\gamma_{\mu}\gamma^{5}\chi
          +\frac{1}{2}\bar{q}\,\sigma^{\mu\nu}q \bar{\chi}\,\sigma_{\mu\nu}\chi\right] \; ,\\
  (\bar{q}\gamma^{5}\chi)(\bar{\chi}\gamma^{5}q) &= -\frac{1}{4}\left[\bar{q}q\bar{\chi}\chi
          -\bar{q}\gamma^{\mu}q\bar{\chi}\gamma_{\mu}\chi
          +\bar{q}\gamma^{5}q\bar{\chi}\gamma^{5}\chi
          +\bar{q}\gamma^{\mu}\gamma^{5}q\bar{\chi}\gamma_{\mu}\gamma^{5}\chi
          +\frac{1}{2}\bar{q}\,\sigma^{\mu\nu}q \bar{\chi}\,\sigma_{\mu\nu}\chi\right] \; ,
\end{align}
\begin{align}
  (\bar{\chi}  q) (\bar{q}  \gamma^5  \chi) &=
    -\dfrac{1}{4} \Bigg[ \bar{q}  q \bar{\chi}  \gamma^5 \chi
    + \bar{q}  \gamma^\mu  q \bar{\chi}  \gamma^\mu \gamma^5  \chi
    + i \epsilon_{\mu \nu \alpha \beta} \bar{q}  \sigma^{\mu \nu}  q \bar{\chi}
                                              \sigma^{\alpha \beta}\chi
    + \bar{q}  \gamma_\mu \gamma^5  q \bar{\chi}  \gamma^\mu \chi
    + \bar{q}  \gamma^5  q \bar{\chi}  \chi  \Bigg] \; ,\\
  (\bar{\chi}  \gamma^5  q) (\bar{q}  \chi) &=
    -\dfrac{1}{4} \Bigg[ \bar{q}  q \bar{\chi}  \gamma^5 \chi
    - \bar{q}  \gamma^\mu  q \bar{\chi}  \gamma^\mu \gamma^5  \chi
    + i \epsilon_{\mu \nu \alpha \beta}  \bar{q}  \sigma^{\mu \nu}  q \bar{\chi}
                                              \sigma^{\alpha \beta}\chi
    - \bar{q}  \gamma_\mu \gamma^5  q \bar{\chi}  \gamma^\mu \chi
    + \bar{q}  \gamma^5  q \bar{\chi}  \chi  \Bigg] \; .
\end{align}

\bibliography{draft_v5}
\end{document}